\documentclass[sn-mathphys,iicol]{sn-jnl}% Math and Physical Sciences Reference Style

\usepackage[normalem]{ulem}
\newcommand{\msout}[1]{\text{\sout{\ensuremath{#1}}}}

\jyear{2022}

\begin{document}

%\title{Grammar-aware implementation of \\
%Quantum Natural Language Processing on IBMQ hardware}

\title[Article Title]{
Grammar-aware sentence classification
on quantum computers}

% \author{Konstantinos Meichanetzidis,
% Alexis Toumi, Giovanni de Felice, Bob Coecke}

\author*{\fnm{Konstantinos} \sur{Meichanetzidis}}\email{k.mei@quantinuum.com}

\author{\fnm{Alexis} \sur{Toumi}}

\author{\fnm{Giovanni} \sur{de Felice}}

\author{\fnm{Bob} \sur{Coecke}}

\affil{Quantinuum, 17 Beaumont Street, Oxford OX1 2NA, United Kingdom}

\affil{Department of Computer Science, University of Oxford, OX1 3QD, United Kingdom}

\abstract{
Natural language processing (NLP) is
at the forefront of great advances in contemporary AI,
and it is arguably one of the most challenging areas of the field.
At the same time, in the area of Quantum Computing (QC), with the steady growth of quantum hardware and notable improvements towards implementations of quantum algorithms, we are approaching an era when quantum computers
perform tasks that cannot be done on classical computers with a reasonable amount of resources.
This provides an new range of opportunities for AI,
and for NLP specifically.
In this work, we work with the Categorical Distributional Compositional (DisCoCat) model of natural language meaning,
whose underlying mathematical underpinnings
make it amenable to quantum instantiations.
Earlier work on fault-tolerant quantum algorithms has already demonstrated potential quantum advantage for NLP, notably employing DisCoCat.
In this work, we focus on the capabilities of 
noisy intermediate-scale quantum (NISQ) hardware
and perform the first implementation of an NLP task on a NISQ processor,
using the DisCoCat framework.
Sentences
are instantiated as parameterised quantum circuits;
word-meanings are embedded in quantum states using parameterised quantum-circuits
and the sentence's grammatical structure
faithfully manifests as a pattern of entangling operations which compose the word-circuits into a sentence-circuit.
The circuits' parameters are trained using a classical optimiser in a supervised NLP task of binary classification.
Our novel QNLP model
shows concrete promise for scalability as the quality of the quantum hardware improves in the near future
and solidifies a novel branch of experimental research at the intersection of QC and AI.
}

\keywords{string diagrams, compositionality, natural language processing, quantum computing}

\maketitle

\section{Introduction}

%NLP
NLP is a rapidly evolving area of AI of both theoretical importance and practical interest \cite{10.5555/555733,BlackburnBos05}.
Large language models,
such as the relatively recent GPT-3 with its 175 billion parameters \cite{brown2020language},
show impressive results on general NLP tasks and one dares to claim that humanity is entering Turing-test territory \cite{10.1093/mind/LIX.236.433}.
Such models, almost always based on neural networks, work by learning to model conditional probability distributions of words in the context of other words.
The textual data on which they are trained are mined from large text corpora; the probability distribution to be learned captures the statistical patterns of cooccurence of words in the data.
Due to this, in this work we shall refer to such models as \emph{distributional}.
%Then, one can sample from the learned distribution, and generate text when given a prompt from the user.

NLP technology becomes increasingly entangled with everyday life as part of search engines, personal assistants, information extraction and data-mining algorithms,
medical diagnoses,
and even bioinformatics \cite{Searls2002,Zeng2015}.
Despite success in both language understanding and language generation,
under the hood of
mainstream NLP models one exclusively
finds deep neural networks,
which famously suffer the criticism of being uninterpretable black boxes \cite{buhrmester2019analysis}.

%NON-BLACKBOX NLP -> DISCOCAT
One way to bring transparency to said black boxes, is to explicitly incorporate \emph{linguistic structure}, such as \emph{grammar} and \emph{syntax} \cite{Lambek58themathematics,MONTAGUE2008,Chomsky1957-CHOSS-2}, into distributional language models.
Note, in this work we will use the terms grammar and syntax interchangeably, the essence of these terms being that they refer to structural information that the textual data could be supplemented with.
A prominent approach attempting this merge is the
\emph{Distributional Compositional Categorical} model of natural language meaning (DisCoCat) \cite{coecke2010mathematical,GrefSadr,KartSadr},
which pioneered the paradigm of combining explicit
grammatical (or syntactic) \emph{structure} with distributional (or statistical) methods for encoding and computing meaning (or semantics).
There has also been follow-up related work on neural-based models where syntax is also incorporated
in a recursive neural network, where the syntactic structures dictate the order of the recursive calls to the recurring cell \cite{socher-etal-2013-recursive}.
This approach also provides the tools
for modelling linguistic phenomena such as lexical entailment and ambiguity,
as well as the transparent construction of syntactic structures like, relative and possessive pronouns \cite{Sadrzadeh2013,Sadrzadeh2014}, conjunction, disjunction, and negation \cite{lewis2020logical}.

From a modern lens,
DisCoCat as it is presented in the literature, is a \emph{tensor network language model}.
Recently,
the motivation for designing interpretable AI systems has caused a surge in the use of tensor networks in language modelling \cite{pestun2017tensor,gallego2019language,bradley2019modeling,efthymiou2019tensornetwork}.
A tensor network is a
graph whose vertices are endowed with tensors.
Every vertex has an arity, ie a number of edges to which it belongs which represent the tensor's indices.
Edges represent tensor contractions, ie identification of the indices joined by the edge and summation over their range (Eistein summation).
Intuitively, a tensor network is a compressed representation of a multiliear map.
Tensor networks, have been used to capture probability distributions of complex many-body systems, both classical and quantum,
and they also have a formal appeal as rigorous algebraic tools \cite{EisertReviewTNS,OrusReviewTNS}.

%QUANTUM COMPUTING
Quantum computing (QC) is a field which, in parallel with NLP is growing at an extremely fast pace.
The prominence of QC is now well-established,
especially after the experiments aiming to demonstrate quantum advantage for the specific task of sampling from random quantum circuits \cite{Arute2019}.
QC has the potential to reach the whole range of human interests,
from foundations of physics and computer science, to applications in engineering, finance, chemistry, and optimisation problems \cite{bharti2021noisy}, and even procedural map generation \cite{Wootton_2020}. % in medicine and the fuel industry [{\bf refs}].

%QML
In the last half-decade, the natural conceptual fusion
of QC with AI, and especially the subfield of AI known as
machine learning (ML), has lead to a plethora of novel and exciting advancements.
The quantum machine learning (QML) literature has reached an immense size considering its young age,
with the cross-fertilisation of ideas and methods between fields of research as well as academia and industry
being a dominant driving force.
The landscape
includes using quantum computers for subroutines
in ML algorithms for executing linear algebra operations \cite{Harrow_2009},
or quantising classical machine learning algorithms based on neural networks \cite{Beer_2020}, support vector machines, clustering \cite{NEURIPS2019_16026d60}, or artificial agents who learn from interacting with their environment \cite{Dunjko_2016},
and even quantum-inspired and dequantised classical algorithms which nevertheless retain a complexity theoretic advantage \cite{Chia_2020}.
Small-scale classification experiments have also been implemented with quantum technology \cite{Havlicek2019,Li_2015}.

%QNLP = QC+NLP
From this collection of ingredients
there organically emerges the interdisciplinary field of
Quantum Natural Language Processing (QNLP),
a research area still in its infancy \cite{Zeng2016,O_Riordan_2020,wiebe2019quantum,bausch2020quantum,chen2002quantum},
combines NLP and QC and seeks 
novel quantum language model designs and quantum algorithms for NLP tasks.
Building on the recently established methodology of QML,
one imports QC algorithms to obtain theoretical speedups for specific NLP tasks
or
use the quantum Hilbert space as a feature space in which NLP tasks are to be executed.

%QUANTUM DISCOCAT AND ADVANTAGE
The first paper on QNLP using the DisCoCat framework by Zeng and Coecke \cite{Zeng2016},
introduced an approach where a standard NLP task are instantiated as quantum computations.
The task of sentence similarity was reduced to the closest-vector problem, for which there exists a quantum algorithm providing a quadratic speedup, albeit assuming a Quantum Random Access Memory (QRAM).
The mapping of the NLP taks to a quantum computation is attributed to the mathematical similarity
of the structures
underlying DisCoCat and quantum theory.
This similarity becomes apparent when both are expressed in the graphical language of string diagrams of monoidal categories or process theories \cite{CKbook}.
The categorical formulation of quantum theory is known as
Categorical Quantum Mechanics (CQM) \cite{1319636}
and it becomes apparent that the string diagrams describing CQM are tensor networks endowed with a graphical language in the form of a rewrite-system (a.k.a. diagrammatic algebra with string diagrams).
The language of string diagrams places syntactic structures and quantum processes on equal footing, and thus allows the canonical instantiation of grammar-aware quantum models for NLP.

%QUANTUM DISCOCAT ON NISQ
In this work, we bring DisCoCat
to the current age of noisy intermediate-scale quantum (NISQ) devices
by performing the first-ever proof-of-concept QNLP experiment on actual quantum processors.
We employ the framework introduced in Ref. \cite{meichanetzidis2020quantum} by
adopting the paradigm of parameterised quantum circuits (PQCs) as quantum machine learning models
\cite{Schuld2020,Benedetti2019},
which currently dominates near-term algorithms.
PQCs can be used to parameterise quantum states and processes, as well as complex probability distributions,
and so they can be used in NISQ machine learning pipelines.
The framework we use in this work allows for the execution of experiments involving non-trivial text corpora,
which moreover involves complex grammatical structures. 
The specific task we showcase here is binary classification of sentences in a supervised-learning hybrid classical-quantum QML setup.

%DISCOCAT SENTENCE
\begin{figure}[t]
\centering
\includegraphics[width=\columnwidth]{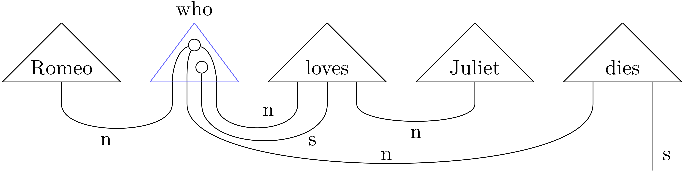}
\caption{Diagram for ``Romeo who loves Juliet dies''.
The grammatical reduction is generated by the nested pattern of non-crossing cups,
which connect words through wires of types $n$ or $s$.
Grammaticality is verified by only one $s$-wire left open.
The diagram represents the meaning of the whole sentence from a process-theoretic point of view.
The relative pronoun `who' is modeled by the Kronecker tensor. % \cite{Sadrzadeh2013}.
Interpreting the diagram in CQM, it represents a quantum state.
}
\label{fig:DisCoCat-sent}
\end{figure}

\section{The model}

%A GRAMMATICAL SENTENCE IS A PROCESS DIAGRAM
% \emph{Sentence Diagrams:}
The DisCoCat model relies on algebraic models of grammar which use types and reduction-rules to mathematically formalise syntactic structures of sentences. Historically, the first formulation of DisCoCat employed pregroup grammars (Appendix \ref{app:pregroups}), which were developed by Lambek \cite{lambekword}. However, \emph{any other typeological grammar}, such as Combinatory Categorial Grammar (CCG), can be used to instantiate DisCoCat models.

In this work we will work with \emph{pregroup grammars}, to stay close to the existing DisCoCat literature.
% Future work involves experimenting with other typeological grammars.
In a pregroup grammar,
a sentence's type is \emph{composed of} a finite product of words $\sigma=\prod_i w_i$.
A parser tags a word $w\in \sigma$ with its part of speech.
Accordingly, $w$ is assigned a \emph{pregroup type} $t_w=\prod_{i} b_i^{\kappa_i}$ comprising a product of \emph{basic (or atomic) types} $b_i$ from the finite set $B$. Each type carries an \emph{adjoint order} $\kappa_i\in\mathbb{Z}$.
Pregroup parsing is efficient; specifically it is linear time under reasonable assumptions \cite{Preller}.
The type of a sentence is the product of the types of its words and it is deemed grammatical iff it type-reduces to the special type $s^0\in B$, i.e. the sentence-type, $t_\sigma = \prod_w t_w \to s^0$.
Reductions are performed by
iteratively applying pairwise annihilations
of basic types
with adjoint orders of the form
$b^{i} b^{i+1}$.
%Consider for example the grammatical sentence ``Romeo who loves Juliet dies''.
%The types assigned to the words of this sentence are as follows.
% Nouns get typed as $t_{\mathrm{Romeo}}=t_{\mathrm{Juliet}}=n^{0}$,
% the transitive verb is given type $t_{\mathrm{loves}}=n^1 s^0 n^{-1}$,
% the intransitive verb is typed $t_{\mathrm{dies}}=n^1 s^0$,
% and the relative pronoun is typed $t_{\mathrm{who}}=n^1 n^0 s^{-1} n^0$.
% The reduction of this sentence is:
% \begin{align}\label{eq:reduction}
% &t_{\mathrm{Romeo\, who\, loves\, Juliet\, dies}}\\\nonumber
% &= t_{\mathrm{Romeo}} t_{\mathrm{who}} t_{\mathrm{loves}} t_{\mathrm{Juliet}} t_{\mathrm{dies}}\\\nonumber
% &=(n^{0}) (n^1 n^0 s^{-1} n^0) (n^1 s^0 n^{-1}) (n^{0}) (n^1 s^0)\\\nonumber
% &\to \msout{n^{0} n^1} n^0 s^{-1} \msout{n^0 n^1} s^0 \msout{n^{-1} n^{0}} n^1 s^0\\\nonumber
% &\to  n^0 \msout{s^{-1}  s^0}  n^1 s^0 \to \msout{n^0  n^1} s^0 \to s^0 .
% \end{align}
As an example, consider the \emph{grammatical reduction}:\\
$t_{\mathrm{Romeo\, who\, loves\, Juliet\, dies}}
= t_{\mathrm{Romeo}}\, t_{\mathrm{who}}\, t_{\mathrm{loves}}\, t_{\mathrm{Juliet}}\, t_{\mathrm{dies}}
=(n^{0}) (n^1 n^0 s^{-1} n^0) (n^1 s^0 n^{-1}) (n^{0}) (n^1 s^0)
\to \msout{n^{0} n^1} n^0 s^{-1} \msout{n^0 n^1} s^0 \msout{n^{-1} n^{0}} n^1 s^0
\to  n^0 \msout{s^{-1}  s^0}  n^1 s^0 \to \msout{n^0  n^1} s^0 \to s^0$.

\begin{figure}[t]
\centering
\includegraphics[width=.9\columnwidth]{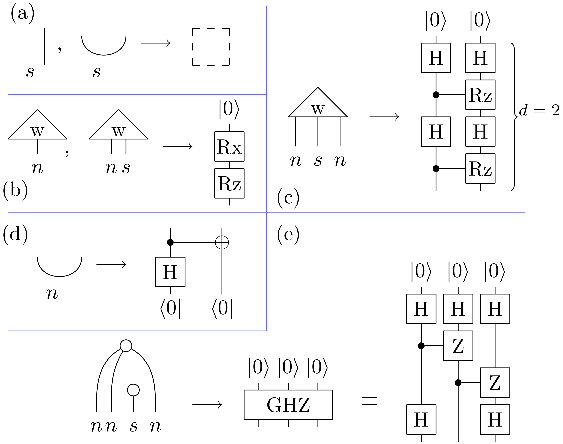}
\caption{
Example instance of mapping
from sentence diagrams to PQCs
where 
$q_n=1$ and $q_s=0$. %, i.e. there are no qubit wires corresponding to the $s$-type in the circuit components.
(a) The dashed square is the empty diagram.
%The parameterised circuit given to each word circuit depends on the word's arity. %, i.e. on how many qubits it is defined.
In this example,
(b) unary word-states are prepared by parameterised $R_x$ rotations followed by $R_z$ rotations
and
(c) $k$-ary word-states are prepared by parameterised word-circuits of width $k$ and depth $d=2$.
(d) The cup is mapped to a Bell effect, i.e.
a CNOT followed by a Hadamard on the control and postselection on $\langle 00\vert$.
(e) The Kronecker tensor modelling the relative pronoun
is mapped to a GHZ state.}
\label{fig:DisCoCat-circ-comps}
\end{figure}

At the core of DisCoCat is a process-theoretic model of natural language meaning.
Process theories are alternatively known as symmetric monoidal (or tensor) categories \cite{baez2009physics}.
Process networks such as those that manifest in DisCoCat can be represented graphically with \emph{string diagrams} \cite{Selinger_2010}.
String diagrams are not just convenient graphical notation,
but they constitute a formal graphical language for reasoning about complex process networks (see Appendix \ref{app:stringdiagrams} for an introduction to string diagrams and concepts relevant to this work).
String diagrams are generated by boxes with input and output wires, with each wire carrying a type.
Boxes can be composed to form process networks by
wiring outputs to inputs and making sure the types are respected.
Output-only processes are called \emph{states}
and input-only processes are called \emph{effects}.
% An effect is the dual of a state.

A grammatical reduction is viewed as a process and so it can be represented as a string diagram.
In string diagrams representing pregroup-grammar reductions,
words are represented as states
and
pairwise type-reductions are represented by a pattern of nested cup-effects (wires bent in a U-shape), and identities (straight wires).
Wires in the string diagram
carry the label of the basic type being reduced.
As an example, in Fig.\ref{fig:DisCoCat-sent} we show the string diagram representing the pregroup reductions
%detailed in Eq.\ref{eq:reduction}
for ``Romeo who loves Juliet dies''.
%Specifically, in Fig.\ref{fig:DisCoCat-circ} there are$n$-type wires and $s$-type wires.
%Every pairwise type-reduction is represented by a cup.
Only the $s$-wire is left open,
which is the \emph{witness of grammaticality}.

% \emph{Sentence Circuits:}

Given a string diagram resulting from the grammatical reduction of a sentence,
we can instantiate a model for natural language processing by giving \emph{semantics} to the string diagram.
This two-step process of constructing a model, where syntax and the semantics are treated separately, is the origin of the framework's name; "Compositional" refers to the string diagram describing structure and "Distributional" refers to the semantic spaces where meaning is encoded and processed.
Any choice of semantics that respects the compositional structure is allowed, and is implemented by component-wise substitution of the boxes and wires in the diagrams.
Such a structure preserving mapping constitutes a \emph{functor},
and ensures that the model instantiation is `canonical'.
Valid choices of semantics range from neural networks, to tensor networks (which was the default choice for the majority of the DisCoCat literature), to quantum processes,
and even hybrid combinations involving components from the whole range of available choice
by interpreting the string diagram correspondingly.

In this work, we focus on the latter choice of \emph{quantum processes},
and in particular, we will be realising quantum processes using \emph{pure quantum circuits}.
As we have described in Ref.\cite{meichanetzidis2020quantum},
the string diagram of the syntactic structure of a sentence $\sigma$ can be canonically mapped
to a PQC $C_\sigma(\theta_\sigma)$
over the parameter set $\theta$.
The key idea here is that
such circuits \emph{inherit their architecture},
in terms of a particular connectivity of entangling gates,
from the
\emph{grammatical reduction} of the sentence.

\begin{figure}[t]
\centering
\includegraphics[width=0.85\columnwidth]{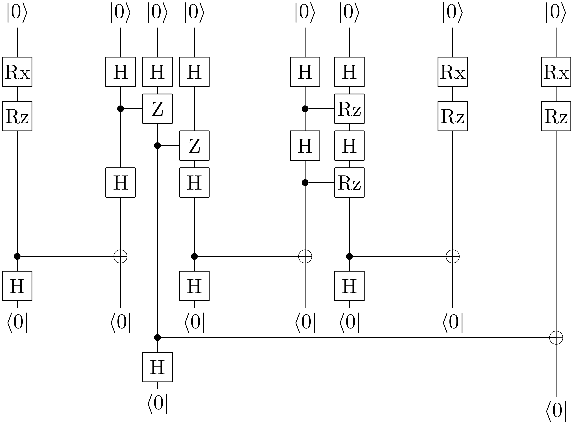}
\caption{The PQC to which ``Romeo who loves Juliet dies'' of Fig.\ref{fig:DisCoCat-sent} is mapped,
with the choices of hyper-parameters of Fig.\ref{fig:DisCoCat-circ-comps}.
As $q_s=0$,
the circuit represents a scalar.
}
\label{fig:DisCoCat-circ}
\end{figure}

Quantum circuits also, being part of pure quantum theory, enjoy a graphical language in terms of string diagrams.
The mapping from sentence diagram to quantum circuit
begins simply by reinterpreting a sentence diagram, such as that of Fig.\ref{fig:DisCoCat-sent}, as a diagram in categorical quantum mechanics (CQM).
The word-state of word $w$ in a sentence diagram
is mapped to a pure quantum state prepared
from a trivial reference product-state by a PQC as
$\vert w(\theta_w)\rangle = C_w(\theta_w) \vert 0 \rangle ^{\otimes q_w}$,
where $q_w=\sum_{b\in t_w} q_b$.
The width of the circuit depends on the number of qubits
assigned to each pregroup type $b\in B$ from which the word-types are composed
and
cups are mapped to Bell effects.

Given a sentence $\sigma$, we instantiate its quantum circuit
by first concatenating in parallel the word-circuits of each word as they appear in the sentence,
corresponding to performing a tensor product,
$C_\sigma(\theta_\sigma) = \bigotimes_w C_w(\theta_w)$
which prepares the state $\vert \sigma(\theta_\sigma)\rangle$ from the all-zeros basis state.
As such, a sentence is parameterised by the concatenation of parameters of its words, $\theta_\sigma = \cup_{w\in\sigma} \theta_w$.
The parameters $\theta_w$ determine the word-embedding $\vert w(\theta_w)\rangle$.
In other words, we use the Hilbert space as a feature space \cite{Havlicek2019,Schuld2019,lloyd2020quantum} in which the word-embeddings are defined.
Finally, we apply Bell effects as dictated by the cup pattern in the grammatical reduction, a function whose result we shall denote $g_\sigma(\vert \sigma(\theta_\sigma)\rangle )$. Note that
in general this procedure prepares an unnormalised quantum state. In the special case where no qubits are assigned to the sentence type, i.e. $q_s=0$, then it is an amplitude which we write as $\langle g_\sigma \vert  \sigma(\theta_\sigma)\rangle $.
Formally, this mapping constitutes a \emph{parameterised functor} from the pregroup grammar category to the category of quantum circuits. The parameterisation is defined via a function from the set of parameters to functors from the aforementioned source and target categories.

% This mapping between sentence diagrams and quantum circuits exists and is well defined due to the monoidal structure
% of the categories describing both pregroup grammars and quantum circuits \cite{Zeng2016,meichanetzidis2020quantum}.

Our model has hyperparameters (Appendix \ref{app:senttocirc}).
The wires of the DisCoCat diagrams we consider
carry types $n$ or $s$.
The number of qubits that we assign to each pregroup type
are $q_n$ and $q_s$.
These determine the \emph{arity} of each word, i.e. the width of the quantum circuit that prepares each word-state.
We set $q_s=0$ throughout this work,
which establishes that the sentence-circuits represent \emph{scalars}.
For a unary word $w$, i.e. a word-state on $1$ qubit,
we choose to prepare using two rotations as
$\mathrm{R}_z(\theta^2_w)R_x(\theta^1_w) \vert 0\rangle$. %, where the first $Z$-rotation can be discarded as it doesn't effect the initial state $\vert0\rangle$.
For a word $w$ of arity $k\geq 2$,
we use a depth-$d$ IQP-style parameterisation \cite{Havlicek2019}
consisting of
$d$ layers where each layer consists of a layer of Hadamard gates followed by a layer of controlled-$Z$ rotations $\mathrm{CR}_z(\theta^i_w)$,
such that $i\in\{1,2,\dots,d(k-1)\}$.
Such circuits are in part motivated by the
conjecture that circuits involving them are classically hard to evaluate \cite{Havlicek2019}.
The relative pronoun ``who'' is mapped to the GHZ circuit,
i.e. the circuit that prepares a GHZ state % from the zero-state
on the number of qubits as determined by $q_n$ and $q_s$.
This is justified by prior work % in the DisCoCat literature
where relative pronouns and other functional words are modelled by a Kronecker tensor (a.k.a. 'spider'), whose entries are all zeros except when all indices are the same for which case the entries are ones \cite{Sadrzadeh2013,Sadrzadeh2014}. It is also known as a 'copy' tensor, as it copies the computational basis.

In Fig.\ref{fig:DisCoCat-circ-comps}
we show an example of choices of word-circuits
for specific numbers of qubits assigned
to each basic pregroup type (Appendix \ref{fig:sent-circ-mapping}).
%In the Appendix we provide further details, in diagrammatic form, on the mapping from sentence diagrams to circuits for any $q_n, q_s, d$.
In Fig.\ref{fig:DisCoCat-circ} we show the corresponding circuit to ``Romeo who loves Juliet dies''.
In practice, we perform the mapping of sentence diagrams to quantum circuits using the Python library {\tt DisCoPy} \cite{felice2020discopy},
which provides a data structure 
for monoidal string-diagrams and enables the instantiation of functors, including functors based on PQCs.

Here, a motivating remark is in order.
In "classical" implementations of DisCoCat,
where the semantics chosen in order to realise a model is in terms of tensor networks,
a sentence diagram represents a vector which results from a tensor contraction.
In this case, meanings of words are encoded in the state-tensors in terms of cooccurrence frequencies or other vector-space word-embeddings \cite{mikolov2013efficient}.
In general, tensor contractions are exponentially expensive to compute.
The cost scales exponentially with the order of the largest tensors present in the tensor network
and the base of the scaling is the dimension of the vector spaces carried by the wires.
However, tensor networks resulting from interpreting syntactic string diagrams over vector spaces and linear maps do not have a generic topology; rather they are tree-like.
This means that contracting tensor networks whose connectivity is given by pregroup reductions are efficiently contractable
as a function of the dimension of the wires carrying the vector spaces playing the role of semantic spaces.
Even in this case however, the dimension of the wires 
for NLP-relevant applications can become prohibitively large (order of hunderds) in practice.
In a fault-tolerant quantum computing setting,
ideally, as is proposed in Ref.\cite{Zeng2016}, one has access to a QRAM and one would be able to efficiently encode such tensor entries as quantum amplitudes using only $\lceil\log_2{d}\rceil$ qubits to encode a $d$-dimensional vector.
However, building a QRAM currently remains challenging \cite{aaronson2015read}.
In the NISQ case, we still attempt to take advantage of the tensor-product structure defined by a collection of qubits which provides an exponentially large Hilbert space as a function of the number of qubits,
and can be used as a feature-space in which the word-embeddings can be trained.
Consequently, we adopt the paradigm of QML in terms of PQCs to carry out near-term QNLP tasks.
Any possible quantum advantage is to be identified heuristically on a case by case basis depending on the task, the data, and the available quantum computing resources.

\section{Classification task}

Now that we have established our construction of
sentence circuits, we describe a simple QNLP task.
The dataset or `labelled corpus' $K=\{(D_\sigma, l_\sigma)\}_\sigma$,
is a finite set of sentence-diagrams
$\{D_\sigma\}_\sigma$ constructed from a finite vocabulary
of words $V$. Each sentence has a binary label $l_\sigma\in\{0,1\}$.
In this work, the labels represent the sentences' truth-values; $0$ for False and $1$ for True.
Our setup trivially generalises to multi-class classification by assigning $q_s=\log_2(\#\text{classes})$ to the $s$-type wire and measuring in the $Z$-basis.

We split $K$ into the training set
$\Delta$
containing the first $\lfloor p \vert \{D_\sigma\}_\sigma \vert  \rceil$
of the sentences,
where $p\in(0,1)$,
and the test set $E$ containing the rest.

% compared to other proof-of-concept supervised QML implementations
% where the data to be classified is generated
% by the process used for classification itself \cite{Havlicek2019}.

We define the \emph{predicted label} as
\begin{equation}
% l^\mathrm{pr}_\sigma(\theta) = \vert\langle00\dots 0\vertC(\theta)_\sigma\vert00\dots 0 \rangle\vert^2 \in [0,1].
l^\mathrm{pr}_\sigma(\theta_\sigma) = \vert \langle g_\sigma \vert \sigma(\theta_\sigma)\rangle \vert ^2 \in [0,1]
\label{eq:predlab}
\end{equation}
from which we can obtain the binary label by rounding to the nearest integer $\lfloor l^\mathrm{pr}_\sigma \rceil \in \{0,1\}$.

The parameters of the words need to be optimised (or trained) so that the predicted labels match the labels in the training set.
The optimiser we invoke is {\tt SPSA} \cite{705889},
a gradient-free optimiser which has shown adequate performance in noisy settings \cite{bonetmonroig2021performance} (Appendix \ref{app:optimisationmethod}).
The \emph{cost function} we define is
\begin{equation}\label{eq:costfunction}
L(\theta) = \sum_{\sigma\in \Delta}
{(l^\mathrm{pr}_\sigma(\theta_\sigma)- l_\sigma)^2  }.
\end{equation}
Minimising the cost function returns the optimal parameters
$\theta^* = \mathrm{argmin} L(\theta)$
from which the model
predicts the labels $l^\mathrm{pr}_\sigma(\theta^*)$.
Essentially, this constitutes \emph{learning a functor} from the grammar category to the category of quantum circuits.
We then quantify the performance
by the training and test errors
$e_\Delta$ and $e_E$, as the proportion of labels predicted incorrectly:
% \begin{align}
% e_\mathrm{tr} &= \sum_{\sigma\in \{C^{tr}_\sigma\}_\sigma}
% {\big\vert\lfloor l^\mathrm{pr}_\sigma(\theta) \rceil - l_\sigma\big\vert}\\
% e_\mathrm{te} &= \sum_{\sigma\in \{C^{te}_\sigma\}_\sigma}
% {\big\vert\lfloor l^\mathrm{pr}_\sigma(\theta) \rceil - l_\sigma\big\vert}
% \end{align}
$$
e_\mathrm{A} = \frac{1}{\vert A\vert} \sum_{\sigma\in {A}}\vert{\big\vert\lfloor l^\mathrm{pr}_\sigma(\theta^*) \rceil - l_\sigma\big\vert}~~,~~~~ A=\Delta, E.
$$
%is the mean Hamming distance between the predicted binary labels and the binary labels in the set $A$.

This supervised learning task of binary classification for sentences
is a special case of \emph{question answering} (QA) \cite{de_Felice_2020,chen2020quantum,zhao2020quantum};
questions are posed as statements and the truth labels are the binary answers.
After training on $\Delta$, the model predicts the answer to a previously unseen question from $E$, which comprises sentences containing words all of which have appeared in $\Delta$.
The optimisation is performed over
the parameters of all the sentences in the training set $\theta = \cup_{\sigma\in\Delta} \theta_\sigma$.
In our experiments, each word appears at least once in the training set and so
$\theta = \cup_{w\in V} \theta_w$.
Note that what is being learned are the inputs, i.e. the quantum word embeddings,
to an entangling process corresponding to the grammar.
Recall that a given sentence-circuit does not necessarily involve the parameters of every word.
However, every word appears in at least one sentence, which
introduces classical correlations between the sentences. This makes such a learning task possible.

In this work, we use artificial data in the form of a very small-scale corpus of grammatical sentences.
We randomly generate sentences using a simple context-free grammar (CFG) whose production rules we define.
Each sentence then is accompanied by its syntax tree by definition of our generation procedure.
The syntax tree can be cast in string-diagram form,
and each CFG-generated sentence-diagram can then be transformed into a DisCoCat diagram (see Appendix \ref{app:cfggen} for details).
% The binary label accompanying a sentence in the corpus
% is given the interpretation of a truth value,
% with $0$ for False and $1$ for True.
Even though the data is synthetic,
we curate the data by assigning labels by hand
so that the truth values among the sentences
are consistent with a story, rendering the classification task semantically non-trivial.

Were one to use real-world datasets of labelled-sentences, one could use a \emph{parser} in order to obtain the syntactic structures of the sentences; the rest of our pipeline would still remain.
Since the writing of this manuscript,
the open-source Python package {\tt lambeq} \cite{kartsaklis2021lambeq} has been made available, which couples to the end-to-end parser {\tt Bobcat}. The parser, given a sentence, returns its syntax tree as a grammatical reduction in the Combinatory Categorial Grammar (CCG). The package also has the capability to translate CCG reductions to pregroup reductions \cite{yeung2021ccgbased}, and so effectively it is the first practical tool introduced for large-scale parsing in pregroup grammar. The string diagrams in {\tt lambeq} are encoded using {\tt DisCoPy}, and thus enables the instantiation of large-scale DisCoCat models on real-world data.

\begin{figure}[t]
\centering
\includegraphics[width=\columnwidth]{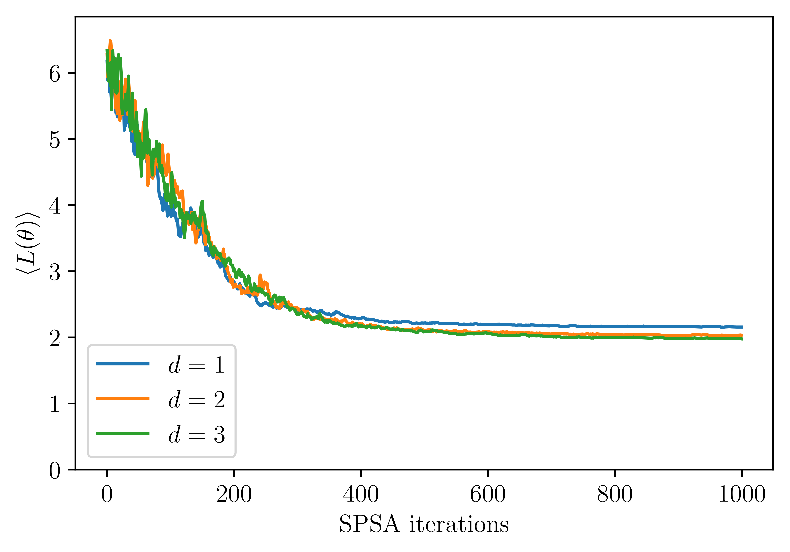}
\includegraphics[width=\columnwidth]{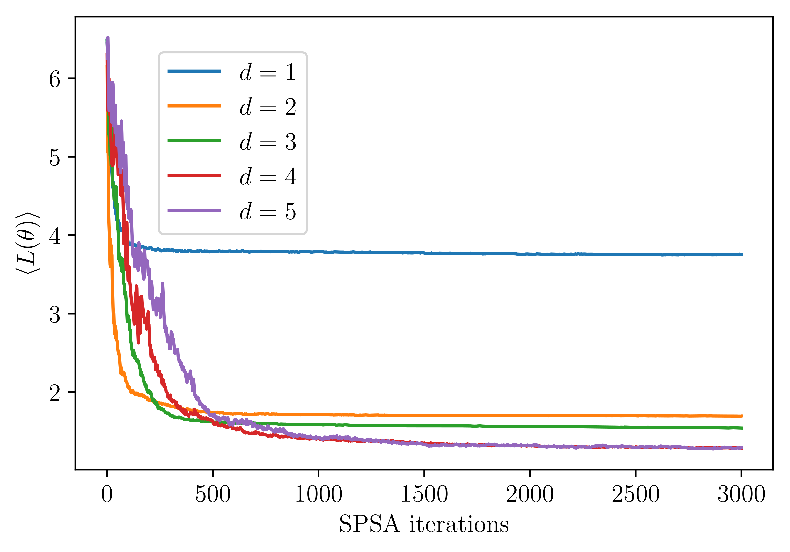}
\begin{picture}(0,0)
\put(-12,230){\includegraphics[scale=0.28]{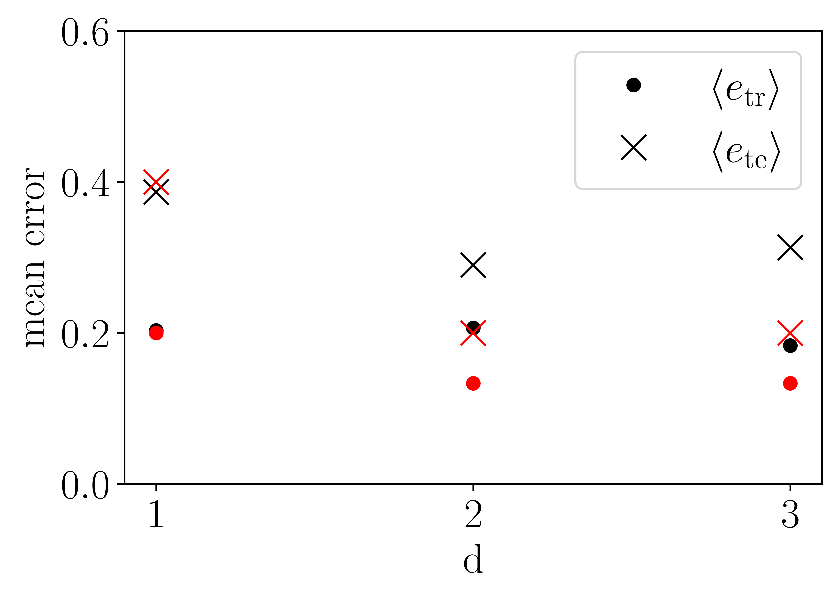} }
\put(-12,78){\includegraphics[scale=0.28]{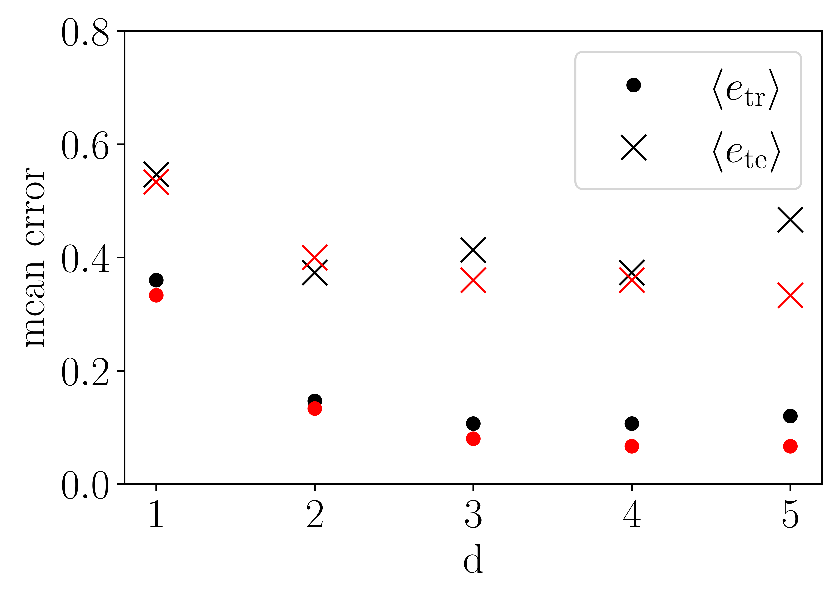}}
\end{picture}
\caption{
Convergence of
mean cost function $\langle L(\theta)\rangle$
vs number of
{\tt SPSA} iterations
for corpus $K_{30}$.
A lower minimum is reached for larger $d$.
(Top) $q_n=1$ and $\vert \theta \vert=8+2d$. % for $d\in\{1,2,3\}$. 
Results are averaged over $20$ realisations.
(Bottom) $q_n=2$ and $\vert \theta\vert=10d$. % for $d\in\{1,2,\dots,5\}$.
Results are averaged over $5$ realisations.
(Insets) Mean training and test errors $\langle e_\mathrm{tr}\rangle$, $\langle e_\mathrm{te}\rangle$ vs $d$.
Using the global optimisation {\tt basinhopping} with local optimisation {\tt Nelder-Mead} (red),
the errors decrease with $d$.}
\label{fig:laptop-convergence}
\end{figure}

\begin{figure}[t]
\centering
\includegraphics[width=\columnwidth]{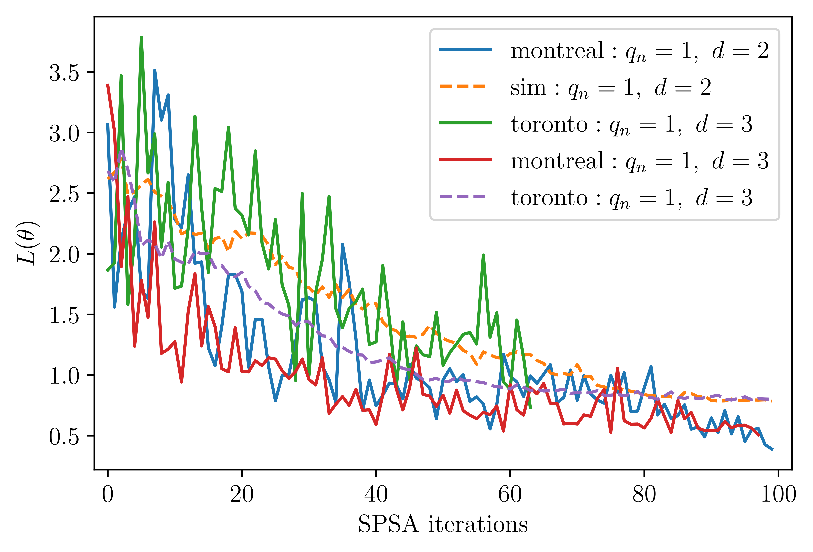}
\caption{Convergence of the cost $L(\theta)$ evaluated on quantum computers 
vs {\tt SPSA} iterations for corpus $K_{16}$.
For
$q_n=1$, $d=2$, for which $\vert\theta \vert =10$,
%$\vert\theta\vert =13$
on {\tt ibmq\_montreal} (blue)
we obtain
$e_\mathrm{tr}=0.125$ and $e_\mathrm{te}=0.5$.
For $q_n=1$, $d=3$, where $\vert\theta \vert =13$,
on {\tt ibmq\_toronto} (green) % and {\tt ibmq\_manhattan}
we get $e_\mathrm{tr}=0.125$
and a lower testing error $e_\mathrm{te}=0.375$.
On {\tt ibmq\_montreal} (red) we get
both lower training and testing errors,
$e_\mathrm{tr}=0$, $e_\mathrm{te}=0.375$ than for $d=2$.
In all cases,
the CNOT-depth of any sentence-circuit after TKET-compilation is at most $3$.
Classical simulations (dashed), averaged over $20$ realisations, agree with behaviour on IBMQ for both cases $d=2$ (yellow) and $d=3$ (purple).
}
\label{fig:ibmq-convergence}
\end{figure}

\subsection{Classical Simulation}

We first show results from classical simulations of the QA task.
The sentence circuits are evaluated exactly
on a classical computer to compute the predicted labels in Eq.\ref{eq:predlab}.
We consider the
corpus $K_{30}$ of $30$ sentences
sampled from the vocabulary of $7$
words (Appendix \ref{app:corpora})
%$V=$\{`Dude', `Walter', `loves', `annoys', `bowls', `abides', `who'\}.
and we set $p=0.5$.
%so that half the sentences are in the training set and half in the test set.
% For unary words we give an Euler decomposition parameterisation.
In Fig.\ref{fig:laptop-convergence} we show the convergence of the cost function,
for $q_n = 1$ and $q_n = 2$,
for increasing word-circuit depth $d$.
To clearly show the decrease in training and test errors as a function of $d$ when invoking the global optimiser {\tt basinhopping} (Appendix \ref{app:optimisationmethod}).

%Note that Rz on a \vert0> is a wasted parameter as it does nothing to the state... So the K_6 experiment is killed.
\subsection{Experiments on IBMQ}

We now turn to readily available NISQ devices
provided by the IBMQ
in order to estimate the predicted labels in Eq.\ref{eq:predlab}.

Before each circuit can be run on a backend,
in this case a superconducting quantum processor,
it first needs to be compiled.
A quantum compiler takes as input a circuit
and a backend and outputs an equivalent circuit
which is compatible with the backend's topology.
A quantum compiler also aims to minimise
the most noisy operations. %, since quantum error correction is not yet available.
For IBMQ, the gate most prone to erros is the entangling CNOT gates.
The compiler we use in this work is TKET \cite{Sivarajah2020}, and
for each circuit-run on a backend, we use the maximum allowed number of shots (Appendix \ref{app:qcompilation}).

% We first consider the small corpus $K_6$ built from $5$ words (see Appendix).
% %from $V=$\{`Romeo', `Juliet', `loves', `dies', `who'\}.
% We set $q_n=1$ and $d=2$,
% and $p=0.8$ so that $5$ sentences are in the training set and the last sentence is kept as test. 
% For unary words we use an $R_z$ rotation.
% %For every evaluation of the cost function under optimisation, the circuits were run on IBMQ quantum computers.
% %We used the {\tt ibmq\_montreal} and {\tt ibmq\_manhattan} backends.
% The parameter space is $5$-dimensional.
% Optimising $L(\theta)$ on {\tt ibmq\_toronto} with {\tt SPSA} for $10$ iterations returns parameter values $\theta^*$ for which
% $e_\Delta=0$.
% The label of the sentence in $E$ is predicted correctly, so trivially $e_E=0$ (see Appendix).

We consider the corpus $K_{16}$ from $6$ words (Appendix \ref{app:corpora}) and set $p=0.5$.
%so that we train over half the corpus and test on the other half.
For every evaluation of the cost function under optimisation, the circuits were run on the IBMQ quantum computers {\tt ibmq\_montreal} and {\tt ibmq\_toronto}.
In Fig.\ref{fig:ibmq-convergence} we show the convergence of the cost function under {\tt SPSA} optimisation
and report the training and testing errors
for different choices of hyperparameters.
This constitutes the first non-trivial QNLP experiment on a programmable quantum processor.
%number of qubits $q_n$ assigned to the $n$-type, IQP-depth $d$, and whether unary words are given $R_z$ or $Euler$ circuits.
% The experiments are performed on
% {\tt ibmq\_montreal} and {\tt ibmq\_toronto}.
%, both of which have quantum volume $32$.
According to Fig.\ref{fig:laptop-convergence},
%the classical simulation of the experiment shows that
scaling up the word-circuits results
in improvement in training and testing errors,
and remarkably,
we observe this on the quantum computer, as well.
%By increasing the word-circuit depth from $d=2$ to $d=3$ we achieve a lower testing error.
This is important for the scalability of our experiment
when future hardware allows for greater circuit sizes
and thus richer quantum-enhanced feature spaces
and grammatically more complex sentences.

% Our QNLP framework is amenable to \emph{quantum advantage}.
% Indeed, in our experiments we made use of postselection
% \emph{On quantum advantage:}
% due to circuit-depth limitations from the noisy quantum hardware.
% Given the ability to perform controlled operations so that the Hadamard test (see Appendix) is within reach,
% we argue that
% for a given choice of parameters for the word-circuits,
% the predicted labels can become hard to predict classically for large enough hyperparameters such as qubits assigned to pregroup types and word-circuit depths.

\section{Discussion and Outlook}

We have performed the first-ever
quantum natural language processing experiment
by means of classification of sentences annotated with binary labels, a special case of QA, on actual quantum hardware.
%We designed a simple proof-of-concept QA task,
%in the `quantum-native' DisCoCat framework
We used a compositional-distributional model of meaning, DisCoCat, constructed
by a structure-preserving mapping
from grammatical reductions of sentences to PQCs.
This proof-of-concept work serves as a demonstration that QNLP is possible on currently available quantum devices and that it is a line of research worth exploring.

A remark on postselection is in order.
QML-based QNLP tasks such as the one implemented in this work
rely on the optimisation of a scalar cost function.
In general, estimating a scalar encoded in an amplitude on a quantum computer requires either postselection or coherent control over arbitrary circuits so that a swap test or a Hadamard test can be performed (Appendix \ref{app:Htest}).
Notably,
in special cases of interest to QML, the Hadamard test can be adapted to NISQ technologies \cite{Mitarai_2019,benedetti2020hardwareefficient}.
In its general form, however, the depth-cost resulting after compilation of controlled circuits becomes prohibitable with current quantum devices.
However, given the rapid improvement in quantum computing hardware,
we envision that such operations will be within reach in the near-term.

Future work includes experimentation with other grammars, such as CCG which returns tree-like diagrams,
and using them to construct PQC-based functors,
as is done in Ref.\cite{socher-etal-2013-recursive}
but with neural networks.
This for example would enable the design of PQC-based functors that do not require postselection,
such as the pregroup-based models where in order for each Bell effect to take place one needs to postselect on measurements involving the qubits on which one wishes to realise a Bell effect.

We also look toward more complex QNLP tasks such as sentence similarity and work with real-world large-scale data using a pregroup parser, as made possible with {\tt lambeq} \cite{kartsaklis2021lambeq}.
In that context, regularisation techniques during training will become important,
which is an increasingly relevant topic for QML that in general deserves more attention \cite{Benedetti2019}.
% This will be required when
% when we move beyond artificial small-scale toy-data and use real-world large-scale data.
%Regularisation methods in NISQy QML are necessary in order to avoid overfitting and poor generalisation ability by the trained model, i.e. exhibit too low training error at the cost of high test error.

In addition,
our DisCoCat-based QNLP framework is naturally generalisable
to accommodate mapping sentences to quantum circuits involving mixed states and quantum channels.
This is useful as mixed states allow for modelling
lexical entailement and ambiguity \cite{piedeleu2015open,bankova2016graded}.
As also stated above, it is possible to define functors in terms of hybrid models where both neural networks and PQCs are involved, where heuristically one aims to quantify the possible advantage of such models compared to strictly classical ones.

Furthermore, note that the word-embeddings are learned in-task in this work.
However, training PQCs to prepare quantum states that serve as word embeddings can be achieved by using the usual NLP objectives \cite{mikolov2013efficient}.
It is interesting to verify that such pretrained word embeddings can be useful in downstream tasks, such as the simple classification task presented in this work.

Finally,
looking beyond the DisCoCat model, it is well-motivated to adopt the
recently introduced DisCoCirc model \cite{coecke2020mathematics} of meaning and its mapping to PQCs \cite{Manifesto},
which allows for QNLP experiments on text-scale real-world data
in a fully-compositional framework.
% In this model, motivated by interpretability in AI,
% word-meanings in DisCoCirc are built bottom-up
% as parameterised states defined on specific tensor factors.
In this model,
nouns are treated as first-class citizen `entities' of a text and makes \emph{sentence composition} explicit.
Entities go through gates which act as modifiers on them,
modelling for example the application of adjectives or verbs.
The model also considers higher-order modifiers, such as adverbs modifying verbs.
This interaction structure, viewed as a process network, can again be used to instantiate models in terms of neural networks, tensor networks, or quantum circuits.
In the latter case, entities are modelled as density matrices carried by wires and their modifiers as quantum channels.

% In this work, however, where we follow the plain DisCoCat model of all words as pure states and grammatical reduction as Bell effects,
% the word-circuit is chosen to be compatible with the limited NISQ architecture and it is the task at hand that determines whether the parameterisation is adequate.

\section*{Acknowledgments}

KM thanks Vojtech Havlicek and Christopher Self for discussions on QML,
Robin Lorenz and Marcello Benedetti for comments on the manuscript,
and the TKET team at Quantinuum
for technical advice on quantum compilation on IBMQ machines.
KM is grateful to the Royal Commission for the Exhibition of 1851
for financial support under a Postdoctoral Research Fellowship.
AT thanks Simon Harrison for financial support through the Wolfson Harrison UK Research
Council Quantum Foundation Scholarship.
We acknowledge the use of IBM Quantum services for this work. The views expressed are those of the authors, and do not reflect the official policy or position of IBM or the IBM Quantum team.\\

{\bf Author contributions} All authors contributed to the theory, the design of the model, and the high-level definition of the classification task.
AT and GDF wrote the DisCoPy library \cite{felice2020discopy} and tested early versions of the experiment on {\tt ibmq\_singapore}.
KM generated and annotated the data, implemented the simulations and experiments, and wrote the manuscript.
BC led the project.\\

{\bf Author information} The authors declare no competing financial interests.\\

{\bf Data availability} The toy datasets of generated sentences used in this work can be found in Appendix \ref{app:corpora}. Further information about the datasets generated and analysed during the current study are available from the corresponding author on reasonable request.

\clearpage

\begin{appendices}

\begin{center}
{\bf \Large Appendix
}
\end{center}

In this supplementary material we begin
by briefly reviewing pregroup grammar.
We then provide the necessary background
to the graphical language of process theories
describe our procedure for generating random sentence diagrams
using a context-free grammar.
For completeness we include the three labelled corpora of sentences we used in this work.
Furthermore, we show details of our mapping
from sentence diagrams to quantum circuits.
Finally we give details on the optimisation methods we used for our supervised quantum machine learning task
and the specific compilation pass we used from CQC's compiler, TKET.

\section{Pregroup Grammar}
\label{app:pregroups}

Pregroup grammars where introduced by Lambek
as an algebraic model for grammar \cite{lambekword}.

A pregroup grammar $G$ is freely generated by
the basic types in a finite set $b\in B$.
Basic types are decorated by an integer $k\in\mathbb{Z}$,
which signifies their adjoint order.
Negative integers $-k$, with $k\in\mathbb{N}$,
are called \emph{left adjoints} of order $k$
and positive integers $k\in\mathbb{N}$ are
called \emph{right adjoints}.
We shall refer to a basic type to some adjoint order (include the zeroth order) simply as `\emph{type}'.
The zeroth order $k=0$ signifies no adjoint action on the basic type and so we often omit it in notation, $b^0=b$.

The pregroup algebra is such that the two kinds of adjoint (left and right) act as left and right inverses under multiplication of basic types
$$ b^{k} b^{k+1} \to \epsilon \to  b^{k+1} b^{k} ,$$
where $\epsilon\in B$ is the trivial or \emph{unit} type.
The left hand side of this reduction is called a \emph{contraction}
and the right hand side an \emph{expansion}.
Pregroup grammar also accommodates \emph{induced steps}
$a \to b$ for $a,b \in B$.
The symbol `$\to$' is to be read as `type-reduction'
and the pregroup grammar sets the rules for which reductions are valid.

Now,
to go from word to sentence,
we consider a finite set of words called the \emph{vocabulary} $V$.
We call the \emph{dictionary} (or lexicon)
the finite set of entries $D \subseteq V \times (B\times\mathbb{Z})^*$.
The star symbol $A^*$
denotes the set of finite strings that can be generated
by the elements of the set $A$.
Each dictionary entry assigns a \emph{product} (or string) of
types to a word $t_w = \prod_i b_i^{k_i}$, $k_i \in\mathbb{Z}$.

Finally,
a pregroup grammar G generates a language $L_G \subseteq V^*$ as follows.
A sentence is a sequence (or list) of words
$\sigma \in V^*$.
The type of a sentence is the product of types of its words
$t_\sigma = \prod_i t_{w_i}$, where $w_i \in V$ and $i\leq \vert\sigma \vert$.
A sentence is \emph{grammatical},
i.e. it belongs to the language generated by the grammar
$\sigma \in L_G$,
if and only if there exists a sequence of reductions so that the type of the sentence reduces to the special \emph{sentence-type} $s\in B$ as
$t_\sigma \to \dots \to s$.
Note that it is in fact possible to type-reduce grammatical sentences \emph{only using contractions}.

\section{String Diagrams}
\label{app:stringdiagrams}

\begin{figure}[h!]
\centering
\includegraphics[width=0.9\columnwidth]{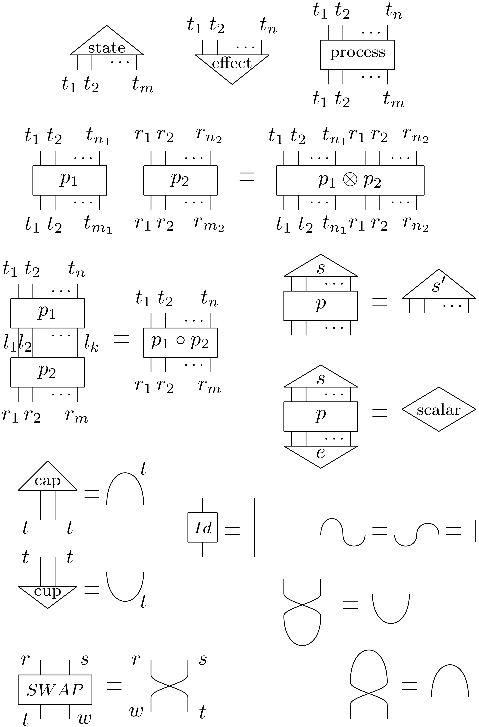}
\caption{Diagrams are read from top to bottom.
States have only outputs, effects have only inputs, processes (boxes) have both input and output wires.
All wires carry types.
Placing boxes side by side is allowed by the monoidal structure
and signifies parallel processes.
Sequential process composition is represented by composing outputs of a box with inputs of another box.
A process transforms a state into a new state.
There are special kinds of states called caps and effects called cups, which satisfy the snake equation which relates them to the identity wire (trivial process).
Process networks freely generated by these generators need not be planar, and so there exists a special process that swaps wires and acts trivially on caps and cups.
}
\label{fig:process-theories}
\end{figure}

String diagrams describing process theories are
generated by states, effects, and processes.
In Fig.\ref{fig:process-theories} we comprehensively show these generators along with constraining equations on them.
String diagrams for process theories
formally describe process networks where only
connectivity matters, i.e. which outputs are connected to which inputs. In other words, the length of the wires carries no meaning and the wires are freely deformable as long as the topology of the network is respected.

It is beyond the purposes of this work to provide a
comprehensive exposition on diagrammatic languages.
We provide the necessary elements which are used for the implementation of our QNLP experiments.

\section{Random Sentence Generation with CFG}
\label{app:cfggen}

A context-free grammar generates a language from a set of production (or rewrite) rules
applied on symbols.
Symbols belong to a finite set $\Sigma$ and
There is a special type $S\in\Sigma$ called initial.
Production rules belong to a finite set $R$
and are of the form $T \to \prod_i T_i$, where $T,T_i\in\Sigma$.
The application of a production rule results in substituting a symbol
with a product (or string) of symbols.
Randomly generating a sentence amounts to starting from $S$
and randomly applying production rules uniformly sampled from the set $R$.
The production ends when all types produced are terminal types,
which are non other than words in the finite vocabulary $V$.

From a process theory point of view, we represent symbols as types
carried by wires.
Production rules are represented as boxes with input and output wires
labelled by the appropriate types.
The process network (or string diagram) describing the production of a sentence
ends with a production rule whose output is the $S$-type.
Then we randomly pick boxes and compose them backwards, always respecting type-matching
when inputs of production rules are fed into outputs of other production rules.
The generation terminates when production rules are applied which have no inputs (i.e. they are states), and they correspond to the words in the finite vocabulary.

\begin{figure}[t]
\centering
\includegraphics[width=\columnwidth]{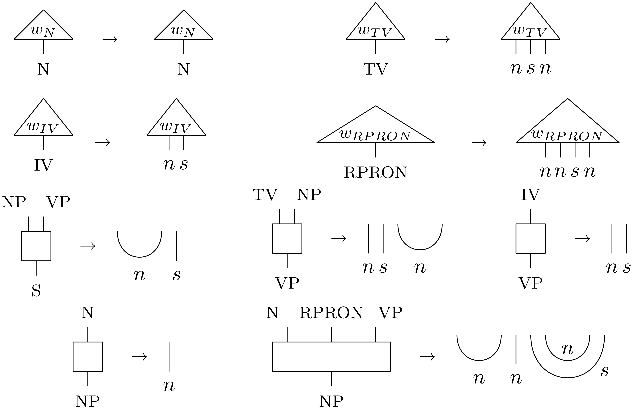}
\caption{CFG generation rules used to produce the corpora $K_{30}, K_6, K_{16}$ used in this work, represented as string-diagram generators,
where $w_N\in V_{N}$, $w_{TV}\in V_{TV}$, $w_{IV}\in V_{IV}$, $w_{RPRON}\in V_{RPRON}$.
They are mapped to pregroup reductions by mapping CFG symbols to pregroup types, and so CFG-states are mapped to DisCoCat word-states and production boxes are mapped to products of cups and identities.
Note that the pregroup unit $\epsilon$ is the empty wire and so it is never drawn.
Pregroup type contractions correspond to cups and expansions to caps.
Since grammatical reduction are achievable only with contractions,
only cups are required for the construction of sentence diagrams.
}
\label{fig:rules-translation}
\end{figure}

In Fig.\ref{fig:rules-translation} (on the left hand side of the arrows) we show the string-diagram generators
we use to randomly produce sentences from a vocabulary of words
composed of nouns, transitive verbs, intransitive verbs, and relative pronouns.
The corresponding types of these parts of speech are $N, TV, IV, RPRON$.
The vocabulary is the union of the words of each type,
$V=V_{N}\cup V_{TV}\cup V_{IV}\cup V_{RPRON}$.

Having randomly generated a sentence from the CFG,
its string diagram can be translated into a pregroup sentence diagram.
To do so we use the translation rules shown in Fig.\ref{fig:rules-translation}.
Note that a cup labeled by the basic type $b$ is used to represent a contraction $b^{k} b^{k+1} \to \epsilon$.
Pregroup grammars are weakly equivalent to context-free grammars,
in the sense that they generate the same language \cite{Buszkowski_pregroupgrammars,287565}.

\section{Corpora}
\label{app:corpora}

Here we present the sentences and their labels used in the experiments presented in the main text.

The types assigned to the words of this sentence are as follows.
Nouns get typed as $t_{w\in V_N}=n^{0}$,
transitive verbs are given type $t_{w\in V_{TV}}=n^1 s^0 n^{-1}$,
intransitive verbs are typed $t_{w\in IV}=n^1 s^0$,
and the relative pronoun is typed $t_{\mathrm{who}}=n^1 n^0 s^{-1} n^0$.

Corpus $K_{30}$ of $30$ labeled sentences from the vocabulary
$V_{N}=\{\text{'Dude', 'Walter'}\}$, $V_{TV}=\{\text{'loves', 'annoys'}\}$, $V_{IV}=\{\text{'abides','bowls'}\}$, $V_{RPRON}=\{\text{'who'}\}$:\\
\noindent [('Dude who loves Walter bowls', 1),\\
 ('Dude bowls', 1),\\
 ('Dude annoys Walter', 0),\\
 ('Walter who abides bowls', 0),\\
 ('Walter loves Walter', 1),\\
 ('Walter annoys Dude', 1),\\
 ('Walter bowls', 1),\\
 ('Walter abides', 0),\\
 ('Dude loves Walter', 1),\\
 ('Dude who bowls abides', 1),\\
 ('Walter who bowls annoys Dude', 1),\\
 ('Dude who bowls bowls', 1),\\
 ('Dude who abides abides', 1),\\
 ('Dude annoys Dude who bowls', 0),\\
 ('Walter annoys Walter', 0),\\
 ('Dude who abides bowls', 1),\\
 ('Walter who abides loves Walter', 0),\\
 ('Walter who bowls bowls', 1),\\
 ('Walter loves Walter who abides', 0),\\
 ('Walter annoys Walter who bowls', 0),\\
 ('Dude abides', 1),\\
 ('Dude loves Walter who bowls', 1),\\
 ('Walter who loves Dude bowls', 1),\\
 ('Dude loves Dude who abides', 1),\\
 ('Walter who abides loves Dude', 0),\\
 ('Dude annoys Dude', 0),\\
 ('Walter who annoys Dude bowls', 1),\\
 ('Walter who annoys Dude abides', 0),\\
 ('Walter loves Dude', 1),\\
 ('Dude who bowls loves Walter', 1)]\\

Corpus $K_{6}$ of $6$ labeled sentences from the vocabulary
$V_{N}=\{\text{'Romeo', 'Juliet'}\}$, $V_{TV}=\{\text{'loves'}\}$, $V_{IV}=\{\text{'dies'}\}$, $V_{RPRON}=\{\text{'who'}\}$:\\
\noindent [('Romeo dies', 1.0),\\
 ('Romeo loves Juliet', 0.0),\\
 ('Juliet who dies dies', 1.0),\\
 ('Romeo loves Romeo', 0.0),\\
 ('Juliet loves Romeo', 0.0),\\
 ('Juliet dies', 1.0)]\\

Corpus $K_{16}$ of $16$ labeled sentences from the vocabulary
$V_{N}=\{\text{'Romeo', 'Juliet'}\}$, $V_{TV}=\{\text{'loves', 'kills'}\}$, $V_{IV}=\{\text{'dies'}\}$, $V_{RPRON}=\{\text{'who'}\}$:\\
\noindent [('Juliet kills Romeo who dies', 0),\\
 ('Juliet dies', 1),\\
 ('Romeo who loves Juliet dies', 1),\\
 ('Romeo dies', 1),\\
 ('Juliet who dies dies', 1),\\
 ('Romeo loves Juliet', 1),\\
 ('Juliet who dies loves Juliet', 0),\\
 ('Romeo kills Juliet who dies', 0),\\
 ('Romeo who kills Romeo dies', 1),\\
 ('Romeo who dies dies', 1),\\
 ('Romeo who loves Romeo dies', 0),\\
 ('Romeo kills Juliet', 0),\\
 ('Romeo who dies kills Romeo', 1),\\
 ('Juliet who dies kills Romeo', 0),\\
 ('Romeo loves Romeo', 0),\\
 ('Romeo who dies kills Juliet', 0)]

\section{Sentence to Circuit mapping}
\label{app:senttocirc}

Quantum theory has formally been shown to be a process theory.
Therefore it enjoys a diagrammatic language in terms of string diagrams.
Specifically, in the context of the quantum circuits we construct in our experiments, we use pure quantum theory.
In the case of pure quantum theory,
processes are unitary operations,
or quantum gates in the context of circuits.
The monoidal structure allowing for parallel processes
is instantiated by the tensor product
and
sequential composition is instantiated by sequential composition of quantum gates.

\begin{figure}[t]
\centering
\includegraphics[width=\columnwidth]{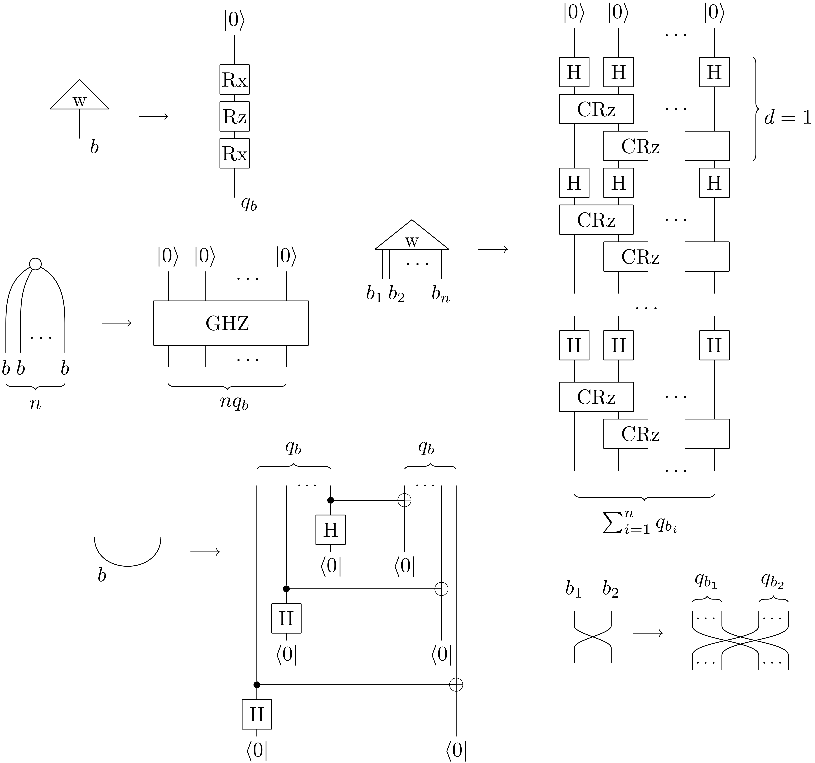}
\caption{Mapping from sentence diagrams to parameterised quantum circuits. Here we show how the generators of sentence diagrams are mapped to generators of circuits, for the hyperparameters we consider in this work.
}
\label{fig:sent-circ-mapping}
\end{figure}

In Fig.\ref{fig:sent-circ-mapping}
we show the generic construction of the mapping
from sentence diagrams to parameterised quantum circuits
for the hyperparameters and parameterised word-circuits we use in this work.

A wire carrying basic pregroup type $b$ is given $q_b$ qubits.
A word-state with only one output wire
becomes a one-qubit-state prepared from $\vert 0\rangle$.
For the preparation of such unary states we choose
the sequence of gates defining an Euler decomposition of one-qubit unitaries
$R_z(\theta_1)\circ R_x(\theta_2) \circ R_z(\theta_3)$.
Word-states with more than one output wires
become multiqubit states on $k>1$ qubits prepared by an IQP-style circuit 
from $\prod_{i=1}^{k}\vert 0\rangle$.
Such a word-circuit is composed of $d$-many layers.
Each layer is composed of a layer of Hadamard gates
followed by a layer in which every neighbouring pair of qubit wires
is connected by a $CR_z(\theta)$ gate,
$\left(\otimes_{i=1}^{k}H\right) \circ \left(\otimes_{i=1}^{k-1} {CR_z}(\theta_i)_{i,i+1}\right)$.
Since all $CR_z$ gates commute with each other it is justified to consider this as a single layer, at least abstractly.
The Kronecker tensor with $n$-many output wires of type $b$
is mapped to a GHZ state on $n q_b$ qubits.
Specifically, GHZ is a circuit that prepares the state
$\sum_{x=0}^{2^{q_b}} \bigotimes_{i=1}^{n} \vert \mathrm{bin}(x)\rangle$,
where $\mathrm{bin}$ is the binary expression of an integer.
The cup of pregroup type $b$
is mapped to $q_b$-many nested Bell effects,
each of which is implemented as a CNOT followed by a Hadamard gate on the control qubit and postselection on $\langle 00 \vert$.

\section{Optimisation Method}
\label{app:optimisationmethod}

The gradient-free otpimisation method we use,
Simultaneous Perturbation Stochastic Approximation ({\tt SPSA}), works as follows.
Start from a random point in parameter space.
At every iteration
pick randomly a direction and \emph{estimate} the derivative
by finite difference with step-size depending on $c$
towards that direction.
This requires two cost function evaluations.
This provides a significant speed up the evaluation of $L(\theta)$.
Then take a step of size depending on $a$ towards
(opposite) that direction
if the derivative has negative (positive) sign.
In our experiments we use {\tt minimizeSPSA} from
the Python package {\tt noisyopt} \cite{noisyopt},
and we set $a=0.1$ and $c=0.1$,
except for the experiment on {\tt ibmq} for $d=3$ for which we set $a=0.05$ and $c=0.05$.

\begin{figure}[t]
\centering
\includegraphics[width=.75\columnwidth]{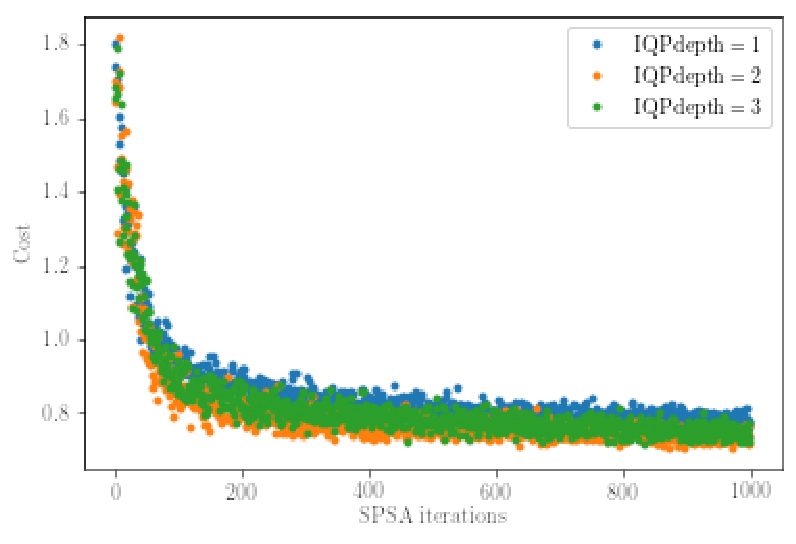}
\caption{Minimisation of binary cross entropy cost function $L^\mathrm{BCE}$ with {\tt SPSA} for the question answering task for corpus $K_{30}$.
}
\label{fig:binaryCE}
\end{figure}

Note that
for classical simulations, we use just-in-time compilation of the cost function
by invoking {\tt jit} from {\tt jax} \cite{jax2018github}.
In addition, the choice of the squares-of-differences cost we defined in Eq.\ref{eq:costfunction}
is not unique.
One can as well use the binary cross entropy
$$L^\mathrm{BCE}(\theta) = -\frac{1}{\vert\Delta \vert}\sum_{\sigma\in\Delta} l_\sigma \log l_\sigma^\mathrm{pr}(\theta_\sigma) + (1-l_\sigma) \log(1-l_\sigma^\mathrm{pr}(\theta_\sigma))$$
and the cost function can be minimised as well,
as shown in Fig.\ref{fig:binaryCE}.

% For our experiment on {\tt ibmq\_manhattan}
% we performed $50$ iterations with {\tt SPSA}
% and then continued for $50$ more after approximately one day.
% As IBMQ's the devices are continually being recalibrated
% and the coherent-noise profile does not stay constant
% under the passage of time periods of the order of magnitude of a day,
% the cost function shows a sharp increase after iteration $50$ in Fig.\ref{fig:ibmq-convergence}.

In our classical simulation of the experiment
we also used {\tt basinhopping} \cite{Olson2012} in combination with
{\tt Nelder-Mead} \cite{NelderMead} from the Python package {\tt SciPy} \cite{scipy}.
{\tt Nelder-Mead} is a gradient-free local optimisation method.
{\tt basinhopping} hops (or jumps) between basins (or local minima) and then returns the minimum over local minima of the cost function,
where each minimum is found by {\tt Nelder-Mead}.
The hop direction is random.
The hop is accepted according to a Metropolis criterion
depending on the the cost function to be minimised
and a temperature.
We used the default temperature value ($1$) and the default number of basin hops ($100$).

\subsection{Error Decay}

\begin{figure}[t]
\centering
\includegraphics[width=0.85\columnwidth]{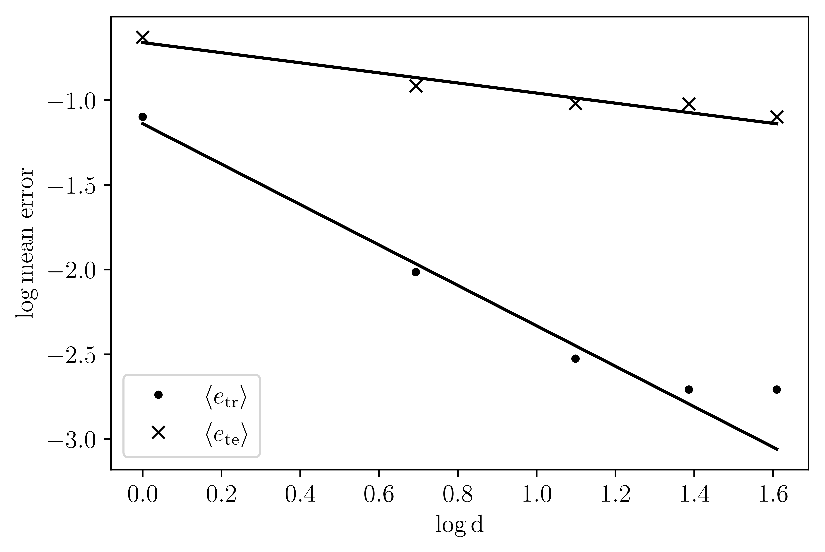}
\caption{Algebraic decay of mean training and testing error for the data displayed in Fig.\ref{fig:laptop-convergence} (bottom) obtained by {\tt basinhopping}. Increasing the depth of the word-circuits results in algebraic decay of the mean training and testing errors. The slopes are $\log e_{\mathrm{tr}}\sim \log -1.2 d$ and $\log e_{\mathrm{te}}\sim -0.3 \log d$.
We attribute the existence of the plateau for $e_\mathrm{tr}$ at large depths is due the small scale of our experiment and the small values for our hyperparameters determining the size of the quantum-enhanced feature space. 
}
\label{fig:error-decay}
\end{figure}

In Fig.\ref{fig:error-decay} we show the
decay of mean training and test errors for the question answering task for corpus $K_{30}$ simulated classically,
which is shown as inset in Fig.\ref{fig:laptop-convergence}.
Plotting in log-log scale we reveal,
at least initially,
an algebraic decay of the errors with the depth of the word-circuits.

% \subsection{Further details for the case of corpus $K_6$}

% Regarding the validity of the smallest experiment we have performed
% on IBMQ using the corpus $K_6$ we do the following.
% To remove the suspicion that the quantum computer
% returns the same labels independently of the parameters given to the circuits and that the task succeeds due to luck,
% we do the following.
% We sample $20$ $\theta$-points in the parameter space and evaluate the Hamming distance
% $e = \sum_{\sigma\in K_{6} }
% {\big\vert\lfloor l^\mathrm{pr}_\sigma(\theta) \rceil - l_\sigma\big\vert}$
% on {\tt ibmq\_manhattan}.
% We find the probabilities $P(e = 0)=0.45$ and $P(e = 3)=0.55$
% which indicate
% that the quantum computer does not always return the correct
% predicted labels for the whole corpus for every point in the parameter space; for almost have the $\theta$-points we sampled we have that $e=3$.

\subsection{On the influence of noise to the cost function landscape}

Regarding optimisation on a quantum computer, we comment on the effect of noise on the optimal parameters.
Consider a successful optimisation of $L(\theta)$
performed on a NISQ device, returning $\theta^*_\mathrm{NISQ}$.
However, if we instantiate the circuits
$C_\sigma(\theta^*_\mathrm{NISQ})$
and evaluate them on a \emph{classical computer}
to obtain the predicted labels $l^\mathrm{CC}_\mathrm{pr}(\theta^*_\mathrm{NISQ})$, we observe that these can in general differ from the labels $l^\mathrm{NISQ}_\mathrm{pr}(\theta^*_\mathrm{NISQ})$ predicted by evaluating the circuits on the \emph{quantum computer}.
In the context of a fault-tolerant quantum computer,
this should not be the case.
However, since there is a non-trivial coherent-noise channel
that our circuits undergo, it is expected that the optimiser's result are affected in this way.

\section{Quantum Compilation}
\label{app:qcompilation}

In order to perform quantum compilation we use ${\tt pytket}$ \cite{Sivarajah2020}.
It is a Python module for interfacing with CQC's TKET, a toolset for quantum programming.
From this toolbox, we need to make use of compilation passes.

At a high level, quantum compilation can be described as follows.
Given a circuit and a device,
quantum operations are decomposed in terms of the devices native gateset. Furthermore, the quantum circuit is reshaped
in order to make it compatible with the device's topology \cite{cowtan_et_al:LIPIcs:2019:10397}.
Specifically,
the compilation pass that we use is 
{\tt default\_compilation\_pass(2)}.
The integer option is set to $2$ for maximum optimisation under compilation \cite{pytket}.

Circuits written in ${\tt pytket}$ can be run on other devices by simply changing the backend being called, regardless whether the hardware might be fundamentally different in terms of what physical systems are used as qubits.
This makes TKET it platform agnostic.
We stress that on IBMQ machines specifically, the native gates are arbitrary single-qubit unitaries (`U3' gate) and entangling controlled-not gates (`CNOT' or `CX').
Importantly, CNOT gates show error rates which are one or even two orders of magnitude larger than error rates of U3 gates.
Therefore, we measure the depth of or circuits in terms of the CNOT-depth.
Using {\tt pytket} this can be obtained by invoking the command {\tt depth\_by\_type(OpType.CX)}.

For both backends used in this work, {\tt ibmq\_montreal} and {\tt ibmq\_toronto}, the reported quantum volume is 32 and the maximum allowed number of shots is $2^{13}$.
%The calibrations for the quantum computers used can be found at {\bf [ref repo]}.

\begin{figure}[t]
\centering
\includegraphics[scale=0.6]{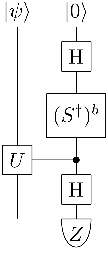}\hspace*{2cm}\includegraphics[scale=0.6]{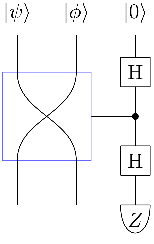}
\caption{
(Left) Circuit for the Hadamard test. Measuring the control qubit in the
computational basis allows one to estimate
$\langle Z \rangle = \mathrm{Re}(\langle \psi\vert U\vert\psi\rangle)$
if $b=0$,
and
$\langle Z \rangle = \mathrm{Im}(\langle \psi\vert U\vert\psi\rangle)$ if $b=1$.
The state $\psi$ can be a multiqubit state, and in this work we are interested in the case $\psi = \vert0\dots 0\rangle$.
(Right) Circuit for the swap test. Sampling from the control qubit allows one to estimate $\langle Z \rangle = \vert\langle \psi \vert \phi \rangle\vert^2$.
}
\label{fig:Htest}
\end{figure}

\begin{figure}[t]
\includegraphics[width=0.9\columnwidth]{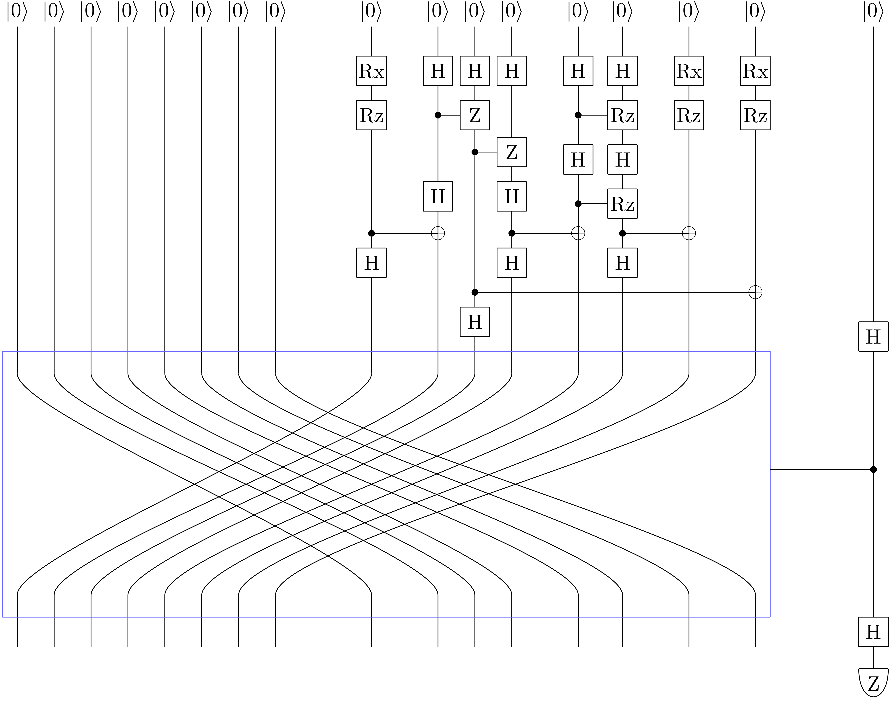}
    \vspace{1cm}

\includegraphics[width=0.9\columnwidth]{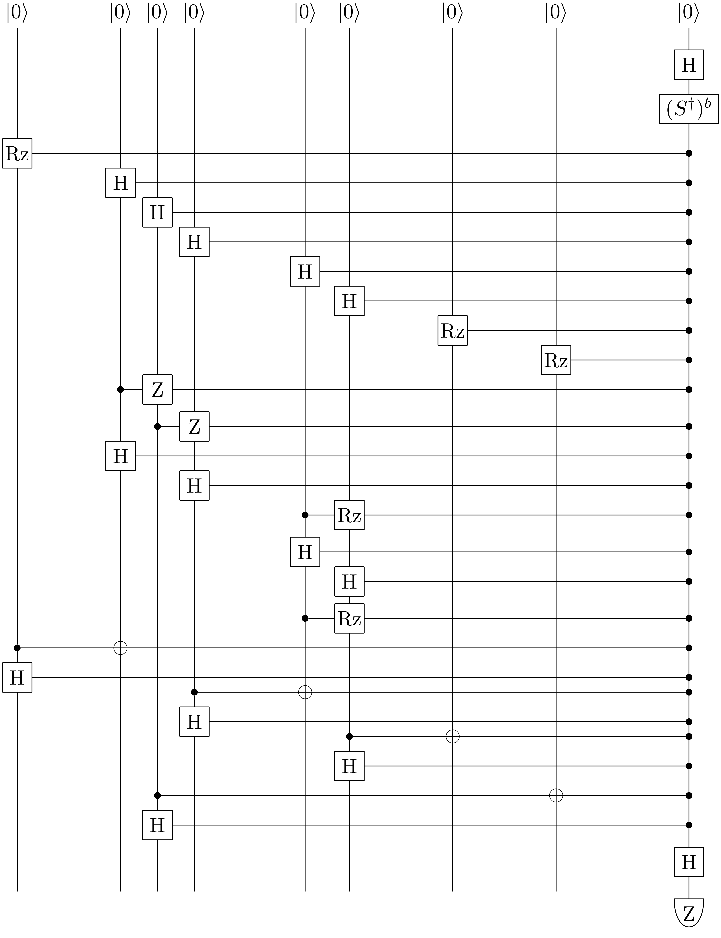}
\caption{ Use of swap test (top) and Hadamard test (bottom) to estimate the norm-squared of the amplitude or the amplitude itself respectively, which is represented by the postselected circuit of Fig.\ref{fig:DisCoCat-circ}.
}
\label{fig:Htest-example}
\end{figure}

\section{Swap Test and Hadamard Test}
\label{app:Htest}

In our binary classification NLP task, the predicted label is the norm squared of
zero-to-zero transition amplitude where the unitary $U$
represents the word-circuits and the circuits that implement the Bell effects as dictated by the grammatical structure.
Estimating $\vert\langle 0\dots 0\vert U \vert0\dots 0\rangle\vert^2$, or the amplitude $\langle 0\dots 0\vert U \vert0\dots 0\rangle$ itself in case one wants to define a cost function where it appears instead of its norm, can be done by postselecting on $\langle 0\dots0\vert$.
However, postselection costs exponential time in the number of postselected qubits;
in our case needs to discard all bitstring sampled from the quantum computer that have Hamming weight other than zero.
This is the procedure we follow in this proof of concept experiment, as we can afford doing so due to the small circuit sizes.

In such a setting, postselection can be avoided by using the swap test to estimate the normed square of the amplitude or the Hadamard test for the amplitude itself \cite{Aharonov2008}.
See Fig.\ref{fig:Htest} for the corresponding circuits of those routines.
In Fig.\ref{fig:Htest-example} we show how the swap test or the Hadamard test can be used to estimate the amplitude represented by the postselected sentence-circuit of Fig.\ref{fig:DisCoCat-circ}.
Furthermore, at least for tasks that are defined such that every sentence corresponds to a circuit, we argue that sentences do not grow arbitrarily long, and so the cost of evaluating the cost function is upper bounded in practical applications.

\end{appendices}

% \bibliographystyle{plain}

% \bibliography{refs}

\begin{thebibliography}{76}
% BibTex style file: bmc-mathphys.bst (version 2.1), 2014-07-24
\ifx \bisbn   \undefined \def \bisbn  #1{ISBN #1}\fi
\ifx \binits  \undefined \def \binits#1{#1}\fi
\ifx \bauthor  \undefined \def \bauthor#1{#1}\fi
\ifx \batitle  \undefined \def \batitle#1{#1}\fi
\ifx \bjtitle  \undefined \def \bjtitle#1{#1}\fi
\ifx \bvolume  \undefined \def \bvolume#1{\textbf{#1}}\fi
\ifx \byear  \undefined \def \byear#1{#1}\fi
\ifx \bissue  \undefined \def \bissue#1{#1}\fi
\ifx \bfpage  \undefined \def \bfpage#1{#1}\fi
\ifx \blpage  \undefined \def \blpage #1{#1}\fi
\ifx \burl  \undefined \def \burl#1{\textsf{#1}}\fi
\ifx \doiurl  \undefined \def \doiurl#1{\url{https://doi.org/#1}}\fi
\ifx \betal  \undefined \def \betal{\textit{et al.}}\fi
\ifx \binstitute  \undefined \def \binstitute#1{#1}\fi
\ifx \binstitutionaled  \undefined \def \binstitutionaled#1{#1}\fi
\ifx \bctitle  \undefined \def \bctitle#1{#1}\fi
\ifx \beditor  \undefined \def \beditor#1{#1}\fi
\ifx \bpublisher  \undefined \def \bpublisher#1{#1}\fi
\ifx \bbtitle  \undefined \def \bbtitle#1{#1}\fi
\ifx \bedition  \undefined \def \bedition#1{#1}\fi
\ifx \bseriesno  \undefined \def \bseriesno#1{#1}\fi
\ifx \blocation  \undefined \def \blocation#1{#1}\fi
\ifx \bsertitle  \undefined \def \bsertitle#1{#1}\fi
\ifx \bsnm \undefined \def \bsnm#1{#1}\fi
\ifx \bsuffix \undefined \def \bsuffix#1{#1}\fi
\ifx \bparticle \undefined \def \bparticle#1{#1}\fi
\ifx \barticle \undefined \def \barticle#1{#1}\fi
\bibcommenthead
\ifx \bconfdate \undefined \def \bconfdate #1{#1}\fi
\ifx \botherref \undefined \def \botherref #1{#1}\fi
\ifx \url \undefined \def \url#1{\textsf{#1}}\fi
\ifx \bchapter \undefined \def \bchapter#1{#1}\fi
\ifx \bbook \undefined \def \bbook#1{#1}\fi
\ifx \bcomment \undefined \def \bcomment#1{#1}\fi
\ifx \oauthor \undefined \def \oauthor#1{#1}\fi
\ifx \citeauthoryear \undefined \def \citeauthoryear#1{#1}\fi
\ifx \endbibitem  \undefined \def \endbibitem {}\fi
\ifx \bconflocation  \undefined \def \bconflocation#1{#1}\fi
\ifx \arxivurl  \undefined \def \arxivurl#1{\textsf{#1}}\fi
\csname PreBibitemsHook\endcsname

%%% 1
\bibitem{10.5555/555733}
\begin{bbook}
\bauthor{\bsnm{Jurafsky}, \binits{D.}},
\bauthor{\bsnm{Martin}, \binits{J.H.}}:
\bbtitle{Speech and Language Processing: An Introduction to Natural Language
  Processing, Computational Linguistics, and Speech Recognition},
\bedition{1st} edn.
\bpublisher{Prentice Hall PTR},
\blocation{USA}
(\byear{2000})
\end{bbook}
\endbibitem

%%% 2
\bibitem{BlackburnBos05}
\begin{bbook}
\bauthor{\bsnm{Blackburn}, \binits{P.}},
\bauthor{\bsnm{Bos}, \binits{J.}}:
\bbtitle{Representation and Inference for Natural Language: A First Course in
  Computational Semantics}.
\bpublisher{Center for the Study of Language and Information},
\blocation{Stanford, CA}
(\byear{2005})
\end{bbook}
\endbibitem

%%% 3
\bibitem{brown2020language}
\begin{botherref}
\oauthor{\bsnm{Brown}, \binits{T.B.}},
\oauthor{\bsnm{Mann}, \binits{B.}},
\oauthor{\bsnm{Ryder}, \binits{N.}},
\oauthor{\bsnm{Subbiah}, \binits{M.}},
\oauthor{\bsnm{Kaplan}, \binits{J.}},
\oauthor{\bsnm{Dhariwal}, \binits{P.}},
\oauthor{\bsnm{Neelakantan}, \binits{A.}},
\oauthor{\bsnm{Shyam}, \binits{P.}},
\oauthor{\bsnm{Sastry}, \binits{G.}},
\oauthor{\bsnm{Askell}, \binits{A.}},
\oauthor{\bsnm{Agarwal}, \binits{S.}},
\oauthor{\bsnm{Herbert-Voss}, \binits{A.}},
\oauthor{\bsnm{Krueger}, \binits{G.}},
\oauthor{\bsnm{Henighan}, \binits{T.}},
\oauthor{\bsnm{Child}, \binits{R.}},
\oauthor{\bsnm{Ramesh}, \binits{A.}},
\oauthor{\bsnm{Ziegler}, \binits{D.M.}},
\oauthor{\bsnm{Wu}, \binits{J.}},
\oauthor{\bsnm{Winter}, \binits{C.}},
\oauthor{\bsnm{Hesse}, \binits{C.}},
\oauthor{\bsnm{Chen}, \binits{M.}},
\oauthor{\bsnm{Sigler}, \binits{E.}},
\oauthor{\bsnm{Litwin}, \binits{M.}},
\oauthor{\bsnm{Gray}, \binits{S.}},
\oauthor{\bsnm{Chess}, \binits{B.}},
\oauthor{\bsnm{Clark}, \binits{J.}},
\oauthor{\bsnm{Berner}, \binits{C.}},
\oauthor{\bsnm{McCandlish}, \binits{S.}},
\oauthor{\bsnm{Radford}, \binits{A.}},
\oauthor{\bsnm{Sutskever}, \binits{I.}},
\oauthor{\bsnm{Amodei}, \binits{D.}}:
Language Models are Few-Shot Learners
(2020)
\end{botherref}
\endbibitem

%%% 4
\bibitem{10.1093/mind/LIX.236.433}
\begin{barticle}
\bauthor{\bsnm{TURING}, \binits{A.M.}}:
\batitle{{I.—COMPUTING MACHINERY AND INTELLIGENCE}}.
\bjtitle{Mind}
\bvolume{LIX}(\bissue{236}),
\bfpage{433}--\blpage{460}
(\byear{1950})
{\href{https://arxiv.org/abs/https://academic.oup.com/mind/article-pdf/LIX/236/433/30123314/lix-236-433.pdf}{{https://academic.oup.com/mind/article-pdf/LIX/236/433/30123314/lix-236-433.pdf}}}.
\doiurl{10.1093/mind/LIX.236.433}
\end{barticle}
\endbibitem

%%% 5
\bibitem{Searls2002}
\begin{barticle}
\bauthor{\bsnm{Searls}, \binits{D.B.}}:
\batitle{The language of genes}.
\bjtitle{Nature}
\bvolume{420}(\bissue{6912}),
\bfpage{211}--\blpage{217}
(\byear{2002}).
\doiurl{10.1038/nature01255}
\end{barticle}
\endbibitem

%%% 6
\bibitem{Zeng2015}
\begin{barticle}
\bauthor{\bsnm{Zeng}, \binits{Z.}},
\bauthor{\bsnm{Shi}, \binits{H.}},
\bauthor{\bsnm{Wu}, \binits{Y.}},
\bauthor{\bsnm{Hong}, \binits{Z.}}:
\batitle{Survey of natural language processing techniques in bioinformatics}.
\bjtitle{Computational and Mathematical Methods in Medicine}
\bvolume{2015},
\bfpage{674296}
(\byear{2015}).
\doiurl{10.1155/2015/674296}
\end{barticle}
\endbibitem

%%% 7
\bibitem{buhrmester2019analysis}
\begin{botherref}
\oauthor{\bsnm{Buhrmester}, \binits{V.}},
\oauthor{\bsnm{Münch}, \binits{D.}},
\oauthor{\bsnm{Arens}, \binits{M.}}:
Analysis of Explainers of Black Box Deep Neural Networks for Computer Vision: A
  Survey
(2019)
\end{botherref}
\endbibitem

%%% 8
\bibitem{Lambek58themathematics}
\begin{botherref}
\oauthor{\bsnm{Lambek}, \binits{J.}}:
The mathematics of sentence structure.
AMERICAN MATHEMATICAL MONTHLY,
154--170
(1958)
\end{botherref}
\endbibitem

%%% 9
\bibitem{MONTAGUE2008}
\begin{barticle}
\bauthor{\bsnm{MONTAGUE}, \binits{R.}}:
\batitle{Universal grammar}.
\bjtitle{Theoria}
\bvolume{36}(\bissue{3}),
\bfpage{373}--\blpage{398}
(\byear{2008}).
\doiurl{10.1111/j.1755-2567.1970.tb00434.x}
\end{barticle}
\endbibitem

%%% 10
\bibitem{Chomsky1957-CHOSS-2}
\begin{bbook}
\bauthor{\bsnm{Chomsky}, \binits{N.}}:
\bbtitle{Syntactic Structures}.
\bpublisher{Mouton}, 
(\byear{1957})
\end{bbook}
\endbibitem

%%% 11
\bibitem{coecke2010mathematical}
\begin{botherref}
\oauthor{\bsnm{Coecke}, \binits{B.}},
\oauthor{\bsnm{Sadrzadeh}, \binits{M.}},
\oauthor{\bsnm{Clark}, \binits{S.}}:
Mathematical Foundations for a Compositional Distributional Model of Meaning
(2010)
\end{botherref}
\endbibitem

%%% 12
\bibitem{GrefSadr}
\begin{bchapter}
\bauthor{\bsnm{Grefenstette}, \binits{E.}},
\bauthor{\bsnm{Sadrzadeh}, \binits{M.}}:
\bctitle{Experimental support for a categorical compositional distributional
  model of meaning}.
In: \bbtitle{The 2014 Conference on Empirical Methods on Natural Language
  Processing.},
pp. \bfpage{1394}--\blpage{1404}
(\byear{2011}).
\bcomment{ar{X}iv:1106.4058}
\end{bchapter}
\endbibitem

%%% 13
\bibitem{KartSadr}
\begin{bchapter}
\bauthor{\bsnm{Kartsaklis}, \binits{D.}},
\bauthor{\bsnm{Sadrzadeh}, \binits{M.}}:
\bctitle{Prior disambiguation of word tensors for constructing sentence
  vectors.}
In: \bbtitle{The 2013 Conference on Empirical Methods on Natural Language
  Processing.},
pp. \bfpage{1590}--\blpage{1601}.
\bpublisher{ACL},
(\byear{2013})
\end{bchapter}
\endbibitem

%%% 14
\bibitem{socher-etal-2013-recursive}
\begin{bchapter}
\bauthor{\bsnm{Socher}, \binits{R.}},
\bauthor{\bsnm{Perelygin}, \binits{A.}},
\bauthor{\bsnm{Wu}, \binits{J.}},
\bauthor{\bsnm{Chuang}, \binits{J.}},
\bauthor{\bsnm{Manning}, \binits{C.D.}},
\bauthor{\bsnm{Ng}, \binits{A.}},
\bauthor{\bsnm{Potts}, \binits{C.}}:
\bctitle{Recursive deep models for semantic compositionality over a sentiment
  treebank}.
In: \bbtitle{Proceedings of the 2013 Conference on Empirical Methods in Natural
  Language Processing},
pp. \bfpage{1631}--\blpage{1642}.
\bpublisher{Association for Computational Linguistics},
\blocation{Seattle, Washington, USA}
(\byear{2013}).
\burl{https://www.aclweb.org/anthology/D13-1170}
\end{bchapter}
\endbibitem

%%% 15
\bibitem{Sadrzadeh2013}
\begin{barticle}
\bauthor{\bsnm{Sadrzadeh}, \binits{M.}},
\bauthor{\bsnm{Clark}, \binits{S.}},
\bauthor{\bsnm{Coecke}, \binits{B.}}:
\batitle{The frobenius anatomy of word meanings i: subject and object relative
  pronouns}.
\bjtitle{Journal of Logic and Computation}
\bvolume{23}(\bissue{6}),
\bfpage{1293}--\blpage{1317}
(\byear{2013}).
\doiurl{10.1093/logcom/ext044}
\end{barticle}
\endbibitem

%%% 16
\bibitem{Sadrzadeh2014}
\begin{barticle}
\bauthor{\bsnm{Sadrzadeh}, \binits{M.}},
\bauthor{\bsnm{Clark}, \binits{S.}},
\bauthor{\bsnm{Coecke}, \binits{B.}}:
\batitle{The frobenius anatomy of word meanings ii: possessive relative
  pronouns}.
\bjtitle{Journal of Logic and Computation}
\bvolume{26}(\bissue{2}),
\bfpage{785}--\blpage{815}
(\byear{2014}).
\doiurl{10.1093/logcom/exu027}
\end{barticle}
\endbibitem

%%% 17
\bibitem{lewis2020logical}
\begin{botherref}
\oauthor{\bsnm{Lewis}, \binits{M.}}:
Towards logical negation for compositional distributional semantics
(2020)
\end{botherref}
\endbibitem

%%% 18
\bibitem{pestun2017tensor}
\begin{botherref}
\oauthor{\bsnm{Pestun}, \binits{V.}},
\oauthor{\bsnm{Vlassopoulos}, \binits{Y.}}:
Tensor network language model
(2017)
\end{botherref}
\endbibitem

%%% 19
\bibitem{gallego2019language}
\begin{botherref}
\oauthor{\bsnm{Gallego}, \binits{A.J.}},
\oauthor{\bsnm{Orus}, \binits{R.}}:
Language Design as Information Renormalization
(2019)
\end{botherref}
\endbibitem

%%% 20
\bibitem{bradley2019modeling}
\begin{botherref}
\oauthor{\bsnm{Bradley}, \binits{T.-D.}},
\oauthor{\bsnm{Stoudenmire}, \binits{E.M.}},
\oauthor{\bsnm{Terilla}, \binits{J.}}:
Modeling Sequences with Quantum States: A Look Under the Hood
(2019)
\end{botherref}
\endbibitem

%%% 21
\bibitem{efthymiou2019tensornetwork}
\begin{botherref}
\oauthor{\bsnm{Efthymiou}, \binits{S.}},
\oauthor{\bsnm{Hidary}, \binits{J.}},
\oauthor{\bsnm{Leichenauer}, \binits{S.}}:
TensorNetwork for Machine Learning
(2019)
\end{botherref}
\endbibitem

%%% 22
\bibitem{EisertReviewTNS}
\begin{botherref}
\oauthor{\bsnm{Eisert}, \binits{J.}}:
Entanglement and tensor network states
(2013)
\end{botherref}
\endbibitem

%%% 23
\bibitem{OrusReviewTNS}
\begin{barticle}
\bauthor{\bsnm{Orús}, \binits{R.}}:
\batitle{Tensor networks for complex quantum systems}.
\bjtitle{Nature Reviews Physics}
\bvolume{1}(\bissue{9}),
\bfpage{538}--\blpage{550}
(\byear{2019}).
\doiurl{10.1038/s42254-019-0086-7}
\end{barticle}
\endbibitem

%%% 24
\bibitem{Arute2019}
\begin{barticle}
\bauthor{\bsnm{Arute}, \binits{F.}},
\bauthor{\bsnm{Arya}, \binits{K.}},
\bauthor{\bsnm{Babbush}, \binits{R.}},
\bauthor{\bsnm{Bacon}, \binits{D.}},
\bauthor{\bsnm{Bardin}, \binits{J.C.}},
\bauthor{\bsnm{Barends}, \binits{R.}},
\bauthor{\bsnm{Biswas}, \binits{R.}},
\bauthor{\bsnm{Boixo}, \binits{S.}},
\bauthor{\bsnm{Brandao}, \binits{F.G.S.L.}},
\bauthor{\bsnm{Buell}, \binits{D.A.}},
\bauthor{\bsnm{Burkett}, \binits{B.}},
\bauthor{\bsnm{Chen}, \binits{Y.}},
\bauthor{\bsnm{Chen}, \binits{Z.}},
\bauthor{\bsnm{Chiaro}, \binits{B.}},
\bauthor{\bsnm{Collins}, \binits{R.}},
\bauthor{\bsnm{Courtney}, \binits{W.}},
\bauthor{\bsnm{Dunsworth}, \binits{A.}},
\bauthor{\bsnm{Farhi}, \binits{E.}},
\bauthor{\bsnm{Foxen}, \binits{B.}},
\bauthor{\bsnm{Fowler}, \binits{A.}},
\bauthor{\bsnm{Gidney}, \binits{C.}},
\bauthor{\bsnm{Giustina}, \binits{M.}},
\bauthor{\bsnm{Graff}, \binits{R.}},
\bauthor{\bsnm{Guerin}, \binits{K.}},
\bauthor{\bsnm{Habegger}, \binits{S.}},
\bauthor{\bsnm{Harrigan}, \binits{M.P.}},
\bauthor{\bsnm{Hartmann}, \binits{M.J.}},
\bauthor{\bsnm{Ho}, \binits{A.}},
\bauthor{\bsnm{Hoffmann}, \binits{M.}},
\bauthor{\bsnm{Huang}, \binits{T.}},
\bauthor{\bsnm{Humble}, \binits{T.S.}},
\bauthor{\bsnm{Isakov}, \binits{S.V.}},
\bauthor{\bsnm{Jeffrey}, \binits{E.}},
\bauthor{\bsnm{Jiang}, \binits{Z.}},
\bauthor{\bsnm{Kafri}, \binits{D.}},
\bauthor{\bsnm{Kechedzhi}, \binits{K.}},
\bauthor{\bsnm{Kelly}, \binits{J.}},
\bauthor{\bsnm{Klimov}, \binits{P.V.}},
\bauthor{\bsnm{Knysh}, \binits{S.}},
\bauthor{\bsnm{Korotkov}, \binits{A.}},
\bauthor{\bsnm{Kostritsa}, \binits{F.}},
\bauthor{\bsnm{Landhuis}, \binits{D.}},
\bauthor{\bsnm{Lindmark}, \binits{M.}},
\bauthor{\bsnm{Lucero}, \binits{E.}},
\bauthor{\bsnm{Lyakh}, \binits{D.}},
\bauthor{\bsnm{Mandr{\`a}}, \binits{S.}},
\bauthor{\bsnm{McClean}, \binits{J.R.}},
\bauthor{\bsnm{McEwen}, \binits{M.}},
\bauthor{\bsnm{Megrant}, \binits{A.}},
\bauthor{\bsnm{Mi}, \binits{X.}},
\bauthor{\bsnm{Michielsen}, \binits{K.}},
\bauthor{\bsnm{Mohseni}, \binits{M.}},
\bauthor{\bsnm{Mutus}, \binits{J.}},
\bauthor{\bsnm{Naaman}, \binits{O.}},
\bauthor{\bsnm{Neeley}, \binits{M.}},
\bauthor{\bsnm{Neill}, \binits{C.}},
\bauthor{\bsnm{Niu}, \binits{M.Y.}},
\bauthor{\bsnm{Ostby}, \binits{E.}},
\bauthor{\bsnm{Petukhov}, \binits{A.}},
\bauthor{\bsnm{Platt}, \binits{J.C.}},
\bauthor{\bsnm{Quintana}, \binits{C.}},
\bauthor{\bsnm{Rieffel}, \binits{E.G.}},
\bauthor{\bsnm{Roushan}, \binits{P.}},
\bauthor{\bsnm{Rubin}, \binits{N.C.}},
\bauthor{\bsnm{Sank}, \binits{D.}},
\bauthor{\bsnm{Satzinger}, \binits{K.J.}},
\bauthor{\bsnm{Smelyanskiy}, \binits{V.}},
\bauthor{\bsnm{Sung}, \binits{K.J.}},
\bauthor{\bsnm{Trevithick}, \binits{M.D.}},
\bauthor{\bsnm{Vainsencher}, \binits{A.}},
\bauthor{\bsnm{Villalonga}, \binits{B.}},
\bauthor{\bsnm{White}, \binits{T.}},
\bauthor{\bsnm{Yao}, \binits{Z.J.}},
\bauthor{\bsnm{Yeh}, \binits{P.}},
\bauthor{\bsnm{Zalcman}, \binits{A.}},
\bauthor{\bsnm{Neven}, \binits{H.}},
\bauthor{\bsnm{Martinis}, \binits{J.M.}}:
\batitle{Quantum supremacy using a programmable superconducting processor}.
\bjtitle{Nature}
\bvolume{574}(\bissue{7779}),
\bfpage{505}--\blpage{510}
(\byear{2019}).
\doiurl{10.1038/s41586-019-1666-5}
\end{barticle}
\endbibitem

%%% 25
\bibitem{bharti2021noisy}
\begin{botherref}
\oauthor{\bsnm{Bharti}, \binits{K.}},
\oauthor{\bsnm{Cervera-Lierta}, \binits{A.}},
\oauthor{\bsnm{Kyaw}, \binits{T.H.}},
\oauthor{\bsnm{Haug}, \binits{T.}},
\oauthor{\bsnm{Alperin-Lea}, \binits{S.}},
\oauthor{\bsnm{Anand}, \binits{A.}},
\oauthor{\bsnm{Degroote}, \binits{M.}},
\oauthor{\bsnm{Heimonen}, \binits{H.}},
\oauthor{\bsnm{Kottmann}, \binits{J.S.}},
\oauthor{\bsnm{Menke}, \binits{T.}},
\oauthor{\bsnm{Mok}, \binits{W.-K.}},
\oauthor{\bsnm{Sim}, \binits{S.}},
\oauthor{\bsnm{Kwek}, \binits{L.-C.}},
\oauthor{\bsnm{Aspuru-Guzik}, \binits{A.}}:
Noisy intermediate-scale quantum (NISQ) algorithms
(2021)
\end{botherref}
\endbibitem

%%% 26
\bibitem{Wootton_2020}
\begin{barticle}
\bauthor{\bsnm{Wootton}, \binits{J.R.}}:
\batitle{Procedural generation using quantum computation}.
\bjtitle{International Conference on the Foundations of Digital Games}
(\byear{2020}).
\doiurl{10.1145/3402942.3409600}
\end{barticle}
\endbibitem

%%% 27
\bibitem{Harrow_2009}
\begin{botherref}
\oauthor{\bsnm{Harrow}, \binits{A.W.}},
\oauthor{\bsnm{Hassidim}, \binits{A.}},
\oauthor{\bsnm{Lloyd}, \binits{S.}}:
Quantum algorithm for linear systems of equations.
Physical Review Letters
\textbf{103}(15)
(2009).
\doiurl{10.1103/physrevlett.103.150502}
\end{botherref}
\endbibitem

%%% 28
\bibitem{Beer_2020}
\begin{botherref}
\oauthor{\bsnm{Beer}, \binits{K.}},
\oauthor{\bsnm{Bondarenko}, \binits{D.}},
\oauthor{\bsnm{Farrelly}, \binits{T.}},
\oauthor{\bsnm{Osborne}, \binits{T.J.}},
\oauthor{\bsnm{Salzmann}, \binits{R.}},
\oauthor{\bsnm{Scheiermann}, \binits{D.}},
\oauthor{\bsnm{Wolf}, \binits{R.}}:
Training deep quantum neural networks.
Nature Communications
\textbf{11}(1)
(2020).
\doiurl{10.1038/s41467-020-14454-2}
\end{botherref}
\endbibitem

%%% 29
\bibitem{NEURIPS2019_16026d60}
\begin{bchapter}
\bauthor{\bsnm{Kerenidis}, \binits{I.}},
\bauthor{\bsnm{Landman}, \binits{J.}},
\bauthor{\bsnm{Luongo}, \binits{A.}},
\bauthor{\bsnm{Prakash}, \binits{A.}}:
\bctitle{q-means: A quantum algorithm for unsupervised machine learning}.
In: \beditor{\bsnm{Wallach}, \binits{H.}},
\beditor{\bsnm{Larochelle}, \binits{H.}},
\beditor{\bsnm{Beygelzimer}, \binits{A.}},
\beditor{\bparticle{d\textquotesingle} \bsnm{Alch\'{e}-Buc}, \binits{F.}},
\beditor{\bsnm{Fox}, \binits{E.}},
\beditor{\bsnm{Garnett}, \binits{R.}} (eds.)
\bbtitle{Advances in Neural Information Processing Systems},
vol. \bseriesno{32},
pp. \bfpage{4134}--\blpage{4144}.
\bpublisher{Curran Associates, Inc.}, 
(\byear{2019}).
\burl{https://proceedings.neurips.cc/paper/2019/file/16026d60ff9b54410b3435b403afd226-Paper.pdf}
\end{bchapter}
\endbibitem

%%% 30
\bibitem{Dunjko_2016}
\begin{botherref}
\oauthor{\bsnm{Dunjko}, \binits{V.}},
\oauthor{\bsnm{Taylor}, \binits{J.M.}},
\oauthor{\bsnm{Briegel}, \binits{H.J.}}:
Quantum-enhanced machine learning.
Physical Review Letters
\textbf{117}(13)
(2016).
\doiurl{10.1103/physrevlett.117.130501}
\end{botherref}
\endbibitem

%%% 31
\bibitem{Chia_2020}
\begin{barticle}
\bauthor{\bsnm{Chia}, \binits{N.-H.}},
\bauthor{\bsnm{Gilyén}, \binits{A.}},
\bauthor{\bsnm{Li}, \binits{T.}},
\bauthor{\bsnm{Lin}, \binits{H.-H.}},
\bauthor{\bsnm{Tang}, \binits{E.}},
\bauthor{\bsnm{Wang}, \binits{C.}}:
\batitle{Sampling-based sublinear low-rank matrix arithmetic framework for
  dequantizing quantum machine learning}.
\bjtitle{Proceedings of the 52nd Annual ACM SIGACT Symposium on Theory of
  Computing}
(\byear{2020}).
\doiurl{10.1145/3357713.3384314}
\end{barticle}
\endbibitem

%%% 32
\bibitem{Havlicek2019}
\begin{barticle}
\bauthor{\bsnm{Havlíček}, \binits{V.}},
\bauthor{\bsnm{Córcoles}, \binits{A.D.}},
\bauthor{\bsnm{Temme}, \binits{K.}},
\bauthor{\bsnm{Harrow}, \binits{A.W.}},
\bauthor{\bsnm{Kandala}, \binits{A.}},
\bauthor{\bsnm{Chow}, \binits{J.M.}},
\bauthor{\bsnm{Gambetta}, \binits{J.M.}}:
\batitle{Supervised learning with quantum-enhanced feature spaces}.
\bjtitle{Nature}
\bvolume{567}(\bissue{7747}),
\bfpage{209}--\blpage{212}
(\byear{2019}).
\doiurl{10.1038/s41586-019-0980-2}
\end{barticle}
\endbibitem

%%% 33
\bibitem{Li_2015}
\begin{botherref}
\oauthor{\bsnm{Li}, \binits{Z.}},
\oauthor{\bsnm{Liu}, \binits{X.}},
\oauthor{\bsnm{Xu}, \binits{N.}},
\oauthor{\bsnm{Du}, \binits{J.}}:
Experimental realization of a quantum support vector machine.
Physical Review Letters
\textbf{114}(14)
(2015).
\doiurl{10.1103/physrevlett.114.140504}
\end{botherref}
\endbibitem

%%% 34
\bibitem{Zeng2016}
\begin{barticle}
\bauthor{\bsnm{Zeng}, \binits{W.}},
\bauthor{\bsnm{Coecke}, \binits{B.}}:
\batitle{Quantum algorithms for compositional natural language processing}.
\bjtitle{Electronic Proceedings in Theoretical Computer Science}
\bvolume{221},
\bfpage{67}--\blpage{75}
(\byear{2016}).
\doiurl{10.4204/eptcs.221.8}
\end{barticle}
\endbibitem

%%% 35
\bibitem{O_Riordan_2020}
\begin{barticle}
\bauthor{\bsnm{O’Riordan}, \binits{L.J.}},
\bauthor{\bsnm{Doyle}, \binits{M.}},
\bauthor{\bsnm{Baruffa}, \binits{F.}},
\bauthor{\bsnm{Kannan}, \binits{V.}}:
\batitle{A hybrid classical-quantum workflow for natural language processing}.
\bjtitle{Machine Learning: Science and Technology}
(\byear{2020}).
\doiurl{10.1088/2632-2153/abbd2e}
\end{barticle}
\endbibitem

%%% 36
\bibitem{wiebe2019quantum}
\begin{botherref}
\oauthor{\bsnm{Wiebe}, \binits{N.}},
\oauthor{\bsnm{Bocharov}, \binits{A.}},
\oauthor{\bsnm{Smolensky}, \binits{P.}},
\oauthor{\bsnm{Troyer}, \binits{M.}},
\oauthor{\bsnm{Svore}, \binits{K.M.}}:
Quantum Language Processing
(2019)
\end{botherref}
\endbibitem

%%% 37
\bibitem{bausch2020quantum}
\begin{botherref}
\oauthor{\bsnm{Bausch}, \binits{J.}},
\oauthor{\bsnm{Subramanian}, \binits{S.}},
\oauthor{\bsnm{Piddock}, \binits{S.}}:
A Quantum Search Decoder for Natural Language Processing
(2020)
\end{botherref}
\endbibitem

%%% 38
\bibitem{chen2002quantum}
\begin{botherref}
\oauthor{\bsnm{Chen}, \binits{J.C.}}:
Quantum computation and natural language processing
(2002)
\end{botherref}
\endbibitem

%%% 39
\bibitem{CKbook}
\begin{bbook}
\bauthor{\bsnm{Coecke}, \binits{B.}},
\bauthor{\bsnm{Kissinger}, \binits{A.}}:
\bbtitle{Picturing Quantum Processes. A First Course in Quantum Theory and
  Diagrammatic Reasoning}.
\bpublisher{Cambridge University Press}, 
(\byear{2017}).
\doiurl{10.1017/9781316219317}
\end{bbook}
\endbibitem

%%% 40
\bibitem{1319636}
\begin{bchapter}
\bauthor{\bsnm{{Abramsky}}, \binits{S.}},
\bauthor{\bsnm{{Coecke}}, \binits{B.}}:
\bctitle{A categorical semantics of quantum protocols}.
In: \bbtitle{Proceedings of the 19th Annual IEEE Symposium on Logic in Computer
  Science, 2004.},
pp. \bfpage{415}--\blpage{425}
(\byear{2004}).
\doiurl{10.1109/LICS.2004.1319636}
\end{bchapter}
\endbibitem

%%% 41
\bibitem{meichanetzidis2020quantum}
\begin{botherref}
\oauthor{\bsnm{Meichanetzidis}, \binits{K.}},
\oauthor{\bsnm{Gogioso}, \binits{S.}},
\oauthor{\bsnm{Felice}, \binits{G.D.}},
\oauthor{\bsnm{Chiappori}, \binits{N.}},
\oauthor{\bsnm{Toumi}, \binits{A.}},
\oauthor{\bsnm{Coecke}, \binits{B.}}:
Quantum Natural Language Processing on Near-Term Quantum Computers
(2020)
\end{botherref}
\endbibitem

%%% 42
\bibitem{Schuld2020}
\begin{botherref}
\oauthor{\bsnm{Schuld}, \binits{M.}},
\oauthor{\bsnm{Bocharov}, \binits{A.}},
\oauthor{\bsnm{Svore}, \binits{K.M.}},
\oauthor{\bsnm{Wiebe}, \binits{N.}}:
Circuit-centric quantum classifiers.
Physical Review A
\textbf{101}(3)
(2020).
\doiurl{10.1103/physreva.101.032308}
\end{botherref}
\endbibitem

%%% 43
\bibitem{Benedetti2019}
\begin{barticle}
\bauthor{\bsnm{Benedetti}, \binits{M.}},
\bauthor{\bsnm{Lloyd}, \binits{E.}},
\bauthor{\bsnm{Sack}, \binits{S.}},
\bauthor{\bsnm{Fiorentini}, \binits{M.}}:
\batitle{Parameterized quantum circuits as machine learning models}.
\bjtitle{Quantum Science and Technology}
\bvolume{4}(\bissue{4}),
\bfpage{043001}
(\byear{2019}).
\doiurl{10.1088/2058-9565/ab4eb5}
\end{barticle}
\endbibitem

%%% 44
\bibitem{lambekword}
\begin{botherref}
\oauthor{\bsnm{Lambek}, \binits{J.}}:
From word to sentence
\end{botherref}
\endbibitem

%%% 45
\bibitem{Preller}
\begin{barticle}
\bauthor{\bsnm{Preller}, \binits{A.}}:
\batitle{Linear processing with pregroups}.
\bjtitle{Studia Logica: An International Journal for Symbolic Logic}
\bvolume{87}(\bissue{2/3}),
\bfpage{171}--\blpage{197}
(\byear{2007})
\end{barticle}
\endbibitem

%%% 46
\bibitem{baez2009physics}
\begin{botherref}
\oauthor{\bsnm{Baez}, \binits{J.C.}},
\oauthor{\bsnm{Stay}, \binits{M.}}:
Physics, Topology, Logic and Computation: A Rosetta Stone
(2009)
\end{botherref}
\endbibitem

%%% 47
\bibitem{Selinger_2010}
\begin{botherref}
\oauthor{\bsnm{Selinger}, \binits{P.}}:
A survey of graphical languages for monoidal categories.
Lecture Notes in Physics,
289--355
(2010).
\doiurl{10.1007/978-3-642-12821-9_4}
\end{botherref}
\endbibitem

%%% 48
\bibitem{Schuld2019}
\begin{botherref}
\oauthor{\bsnm{Schuld}, \binits{M.}},
\oauthor{\bsnm{Killoran}, \binits{N.}}:
Quantum machine learning in feature hilbert spaces.
Physical Review Letters
\textbf{122}(4)
(2019).
\doiurl{10.1103/physrevlett.122.040504}
\end{botherref}
\endbibitem

%%% 49
\bibitem{lloyd2020quantum}
\begin{botherref}
\oauthor{\bsnm{Lloyd}, \binits{S.}},
\oauthor{\bsnm{Schuld}, \binits{M.}},
\oauthor{\bsnm{Ijaz}, \binits{A.}},
\oauthor{\bsnm{Izaac}, \binits{J.}},
\oauthor{\bsnm{Killoran}, \binits{N.}}:
Quantum embeddings for machine learning
(2020)
\end{botherref}
\endbibitem

%%% 50
\bibitem{felice2020discopy}
\begin{botherref}
\oauthor{\bparticle{de} \bsnm{Felice}, \binits{G.}},
\oauthor{\bsnm{Toumi}, \binits{A.}},
\oauthor{\bsnm{Coecke}, \binits{B.}}:
DisCoPy: Monoidal Categories in Python
(2020)
\end{botherref}
\endbibitem

%%% 51
\bibitem{mikolov2013efficient}
\begin{botherref}
\oauthor{\bsnm{Mikolov}, \binits{T.}},
\oauthor{\bsnm{Chen}, \binits{K.}},
\oauthor{\bsnm{Corrado}, \binits{G.}},
\oauthor{\bsnm{Dean}, \binits{J.}}:
Efficient Estimation of Word Representations in Vector Space
(2013)
\end{botherref}
\endbibitem

%%% 52
\bibitem{aaronson2015read}
\begin{barticle}
\bauthor{\bsnm{Aaronson}, \binits{S.}}:
\batitle{Read the fine print}.
\bjtitle{Nature Physics}
\bvolume{11}(\bissue{4}),
\bfpage{291}--\blpage{293}
(\byear{2015})
\end{barticle}
\endbibitem

%%% 53
\bibitem{kartsaklis2021lambeq}
\begin{botherref}
\oauthor{\bsnm{Kartsaklis}, \binits{D.}},
\oauthor{\bsnm{Fan}, \binits{I.}},
\oauthor{\bsnm{Yeung}, \binits{R.}},
\oauthor{\bsnm{Pearson}, \binits{A.}},
\oauthor{\bsnm{Lorenz}, \binits{R.}},
\oauthor{\bsnm{Toumi}, \binits{A.}},
\oauthor{\bparticle{de} \bsnm{Felice}, \binits{G.}},
\oauthor{\bsnm{Meichanetzidis}, \binits{K.}},
\oauthor{\bsnm{Clark}, \binits{S.}},
\oauthor{\bsnm{Coecke}, \binits{B.}}:
lambeq: An Efficient High-Level Python Library for Quantum NLP
(2021)
\end{botherref}
\endbibitem

%%% 54
\bibitem{yeung2021ccgbased}
\begin{botherref}
\oauthor{\bsnm{Yeung}, \binits{R.}},
\oauthor{\bsnm{Kartsaklis}, \binits{D.}}:
A CCG-Based Version of the DisCoCat Framework
(2021)
\end{botherref}
\endbibitem

%%% 55
\bibitem{705889}
\begin{barticle}
\bauthor{\bsnm{{Spall}}, \binits{J.C.}}:
\batitle{Implementation of the simultaneous perturbation algorithm for
  stochastic optimization}.
\bjtitle{IEEE Transactions on Aerospace and Electronic Systems}
\bvolume{34}(\bissue{3}),
\bfpage{817}--\blpage{823}
(\byear{1998}).
\doiurl{10.1109/7.705889}
\end{barticle}
\endbibitem

%%% 56
\bibitem{bonetmonroig2021performance}
\begin{botherref}
\oauthor{\bsnm{Bonet-Monroig}, \binits{X.}},
\oauthor{\bsnm{Wang}, \binits{H.}},
\oauthor{\bsnm{Vermetten}, \binits{D.}},
\oauthor{\bsnm{Senjean}, \binits{B.}},
\oauthor{\bsnm{Moussa}, \binits{C.}},
\oauthor{\bsnm{Bäck}, \binits{T.}},
\oauthor{\bsnm{Dunjko}, \binits{V.}},
\oauthor{\bsnm{O'Brien}, \binits{T.E.}}:
Performance comparison of optimization methods on variational quantum
  algorithms
(2021)
\end{botherref}
\endbibitem

%%% 57
\bibitem{de_Felice_2020}
\begin{barticle}
\bauthor{\bparticle{de} \bsnm{Felice}, \binits{G.}},
\bauthor{\bsnm{Meichanetzidis}, \binits{K.}},
\bauthor{\bsnm{Toumi}, \binits{A.}}:
\batitle{Functorial question answering}.
\bjtitle{Electronic Proceedings in Theoretical Computer Science}
\bvolume{323},
\bfpage{84}--\blpage{94}
(\byear{2020}).
\doiurl{10.4204/eptcs.323.6}
\end{barticle}
\endbibitem

%%% 58
\bibitem{chen2020quantum}
\begin{botherref}
\oauthor{\bsnm{Chen}, \binits{Y.}},
\oauthor{\bsnm{Pan}, \binits{Y.}},
\oauthor{\bsnm{Dong}, \binits{D.}}:
Quantum Language Model with Entanglement Embedding for Question Answering
(2020)
\end{botherref}
\endbibitem

%%% 59
\bibitem{zhao2020quantum}
\begin{barticle}
\bauthor{\bsnm{Zhao}, \binits{Q.}},
\bauthor{\bsnm{Hou}, \binits{C.}},
\bauthor{\bsnm{Liu}, \binits{C.}},
\bauthor{\bsnm{Zhang}, \binits{P.}},
\bauthor{\bsnm{Xu}, \binits{R.}}:
\batitle{A quantum expectation value based language model with application to
  question answering}.
\bjtitle{Entropy}
\bvolume{22}(\bissue{5}),
\bfpage{533}
(\byear{2020})
\end{barticle}
\endbibitem

%%% 60
\bibitem{Sivarajah2020}
\begin{barticle}
\bauthor{\bsnm{Sivarajah}, \binits{S.}},
\bauthor{\bsnm{Dilkes}, \binits{S.}},
\bauthor{\bsnm{Cowtan}, \binits{A.}},
\bauthor{\bsnm{Simmons}, \binits{W.}},
\bauthor{\bsnm{Edgington}, \binits{A.}},
\bauthor{\bsnm{Duncan}, \binits{R.}}:
\batitle{Tket: A retargetable compiler for nisq devices}.
\bjtitle{Quantum Science and Technology}
(\byear{2020}).
\doiurl{10.1088/2058-9565/ab8e92}
\end{barticle}
\endbibitem

%%% 61
\bibitem{Mitarai_2019}
\begin{botherref}
\oauthor{\bsnm{Mitarai}, \binits{K.}},
\oauthor{\bsnm{Fujii}, \binits{K.}}:
Methodology for replacing indirect measurements with direct measurements.
Physical Review Research
\textbf{1}(1)
(2019).
\doiurl{10.1103/physrevresearch.1.013006}
\end{botherref}
\endbibitem

%%% 62
\bibitem{benedetti2020hardwareefficient}
\begin{botherref}
\oauthor{\bsnm{Benedetti}, \binits{M.}},
\oauthor{\bsnm{Fiorentini}, \binits{M.}},
\oauthor{\bsnm{Lubasch}, \binits{M.}}:
Hardware-efficient variational quantum algorithms for time evolution
(2020)
\end{botherref}
\endbibitem

%%% 63
\bibitem{piedeleu2015open}
\begin{botherref}
\oauthor{\bsnm{Piedeleu}, \binits{R.}},
\oauthor{\bsnm{Kartsaklis}, \binits{D.}},
\oauthor{\bsnm{Coecke}, \binits{B.}},
\oauthor{\bsnm{Sadrzadeh}, \binits{M.}}:
Open System Categorical Quantum Semantics in Natural Language Processing
(2015)
\end{botherref}
\endbibitem

%%% 64
\bibitem{bankova2016graded}
\begin{botherref}
\oauthor{\bsnm{Bankova}, \binits{D.}},
\oauthor{\bsnm{Coecke}, \binits{B.}},
\oauthor{\bsnm{Lewis}, \binits{M.}},
\oauthor{\bsnm{Marsden}, \binits{D.}}:
Graded Entailment for Compositional Distributional Semantics
(2016)
\end{botherref}
\endbibitem

%%% 65
\bibitem{coecke2020mathematics}
\begin{botherref}
\oauthor{\bsnm{Coecke}, \binits{B.}}:
The Mathematics of Text Structure
(2020)
\end{botherref}
\endbibitem

%%% 66
\bibitem{Manifesto}
\begin{botherref}
\oauthor{\bsnm{Coecke}, \binits{B.}},
\oauthor{\bparticle{de} \bsnm{Felice}, \binits{G.}},
\oauthor{\bsnm{Meichanetzidis}, \binits{K.}},
\oauthor{\bsnm{Toumi}, \binits{A.}}:
Foundations for Near-Term Quantum Natural Language Processing
(2020)
\end{botherref}
\endbibitem

%%% 67
\bibitem{Buszkowski_pregroupgrammars}
\begin{botherref}
\oauthor{\bsnm{Buszkowski}, \binits{W.}},
\oauthor{\bsnm{Moroz}, \binits{K.}}:
Pregroup Grammars and Context-free Grammars
\end{botherref}
\endbibitem

%%% 68
\bibitem{287565}
\begin{bchapter}
\bauthor{\bsnm{{Pentus}}, \binits{M.}}:
\bctitle{Lambek grammars are context free}.
In: \bbtitle{[1993] Proceedings Eighth Annual IEEE Symposium on Logic in
  Computer Science},
pp. \bfpage{429}--\blpage{433}
(\byear{1993}).
\doiurl{10.1109/LICS.1993.287565}
\end{bchapter}
\endbibitem

%%% 69
\bibitem{noisyopt}
\begin{botherref}
\url{https://github.com/andim/noisyopt}
\end{botherref}
\endbibitem

%%% 70
\bibitem{jax2018github}
\begin{botherref}
\oauthor{\bsnm{Bradbury}, \binits{J.}},
\oauthor{\bsnm{Frostig}, \binits{R.}},
\oauthor{\bsnm{Hawkins}, \binits{P.}},
\oauthor{\bsnm{Johnson}, \binits{M.J.}},
\oauthor{\bsnm{Leary}, \binits{C.}},
\oauthor{\bsnm{Maclaurin}, \binits{D.}},
\oauthor{\bsnm{Wanderman-Milne}, \binits{S.}}:
{JAX}: Composable Transformations of {P}ython+{N}um{P}y programs.
\url{http://github.com/google/jax}
\end{botherref}
\endbibitem

%%% 71
\bibitem{Olson2012}
\begin{barticle}
\bauthor{\bsnm{Olson}, \binits{B.}},
\bauthor{\bsnm{Hashmi}, \binits{I.}},
\bauthor{\bsnm{Molloy}, \binits{K.}},
\bauthor{\bsnm{Shehu}, \binits{A.}}:
\batitle{Basin hopping as a general and versatile optimization framework for
  the characterization of biological macromolecules}.
\bjtitle{Advances in Artificial Intelligence}
\bvolume{2012},
\bfpage{1}--\blpage{19}
(\byear{2012}).
\doiurl{10.1155/2012/674832}
\end{barticle}
\endbibitem

%%% 72
\bibitem{NelderMead}
\begin{barticle}
\bauthor{\bsnm{Gao}, \binits{F.}},
\bauthor{\bsnm{Han}, \binits{L.}}:
\batitle{Implementing the nelder-mead simplex algorithm with~adaptive
  parameters}.
\bjtitle{Computational Optimization and Applications}
\bvolume{51}(\bissue{1}),
\bfpage{259}--\blpage{277}
(\byear{2010}).
\doiurl{10.1007/s10589-010-9329-3}
\end{barticle}
\endbibitem

%%% 73
\bibitem{scipy}
\begin{botherref}
\url{https://pypi.org/project/scipy}
\end{botherref}
\endbibitem

%%% 74
\bibitem{cowtan_et_al:LIPIcs:2019:10397}
\begin{bchapter}
\bauthor{\bsnm{Cowtan}, \binits{A.}},
\bauthor{\bsnm{Dilkes}, \binits{S.}},
\bauthor{\bsnm{Duncan}, \binits{R.}},
\bauthor{\bsnm{Krajenbrink}, \binits{A.}},
\bauthor{\bsnm{Simmons}, \binits{W.}},
\bauthor{\bsnm{Sivarajah}, \binits{S.}}:
\bctitle{{On the Qubit Routing Problem}}.
In: \beditor{\bparticle{van} \bsnm{Dam}, \binits{W.}},
\beditor{\bsnm{Mancinska}, \binits{L.}} (eds.)
\bbtitle{14th Conference on the Theory of Quantum Computation, Communication
  and Cryptography (TQC 2019)}.
\bsertitle{Leibniz International Proceedings in Informatics (LIPIcs)},
vol. \bseriesno{135},
pp. \bfpage{5}--\blpage{1532}.
\bpublisher{Schloss Dagstuhl--Leibniz-Zentrum fuer Informatik},
\blocation{Dagstuhl, Germany}
(\byear{2019}).
\doiurl{10.4230/LIPIcs.TQC.2019.5}.
\burl{http://drops.dagstuhl.de/opus/volltexte/2019/10397}
\end{bchapter}
\endbibitem

%%% 75
\bibitem{pytket}
\begin{botherref}
\url{https://github.com/CQCL/pytket}
\end{botherref}
\endbibitem

%%% 76
\bibitem{Aharonov2008}
\begin{barticle}
\bauthor{\bsnm{Aharonov}, \binits{D.}},
\bauthor{\bsnm{Jones}, \binits{V.}},
\bauthor{\bsnm{Landau}, \binits{Z.}}:
\batitle{A polynomial quantum algorithm for approximating the jones
  polynomial}.
\bjtitle{Algorithmica}
\bvolume{55}(\bissue{3}),
\bfpage{395}--\blpage{421}
(\byear{2008}).
\doiurl{10.1007/s00453-008-9168-0}
\end{barticle}
\endbibitem

\end{thebibliography}

%% BioMed_Central_Bib_Style_v1.01

\end{document}